\documentclass[journal,draftclsnofoot,12pt,onecolumn]{IEEEtran}

\usepackage[nospace,noadjust]{cite}
\usepackage{graphicx}
\usepackage[cmex10]{amsmath}
\usepackage{bbm}
\usepackage{algorithmicx}
\usepackage{algpseudocode}
\usepackage{algorithm}
\usepackage{dsfont}
\usepackage{color}
\usepackage{float}
\usepackage[section]{placeins}

\ifCLASSOPTIONcompsoc
  \usepackage[caption=false,font=normalsize,labelfont=sf,textfont=sf]{subfig}
\else
  \usepackage[caption=false,font=footnotesize]{subfig}
\fi

\makeatletter
\newenvironment{varsubequations}[1]
 {%
  \addtocounter{equation}{-1}%
  \begin{subequations}
  \renewcommand{\theparentequation}{#1}%
  \def\@currentlabel{#1}%
 }
 {%
  \end{subequations}
 }
\makeatother

\newtheorem{theorem}{Theorem}

\algnewcommand\algorithmicinput{\textbf{INPUT:}}
\algnewcommand\INPUT{\item[\algorithmicinput]}

\algnewcommand\algorithmicoutput{\textbf{OUTPUT:}}
\algnewcommand\OUTPUT{\item[\algorithmicoutput]}

\algnewcommand\algorithmicinit{\textbf{Initialization:}}
\algnewcommand\Init{\item[\algorithmicinit]}
\newcommand{\BZ}{\textcolor{black}}

\newcommand{\br}{\boldsymbol{r}}

\newcommand{\bx}{\boldsymbol{x}}
\newcommand{\by}{\boldsymbol{y}}
\newcommand{\bz}{\boldsymbol{z}}
\newcommand{\NU}{\mathcal{U}}
\newcommand{\NA}{\mathcal{A}}
\newcommand{\NN}{\mathcal{N}}

\newcommand{\Ncal}{\mathcal{N}}
\newcommand{\Kcal}{\mathcal{K}}
\begin{document}
\title{Scalable Spectrum Allocation and User Association in Networks with Many Small Cells}
\author{Binnan Zhuang, Dongning Guo, Ermin Wei and Michael L. Honig\\
\hfil\\
\thanks{
B.~Zhuang was with the Department of Electrical
  Engineering and Computer Science at Northwestern University,
  Evanston, IL, 60208, USA.
  He is now with Samsung Semiconductor, Inc., Modem R\&D Lab, San Diego, CA, USA.
D.~Guo, E.~Wei and M.~L.~Honig are with the Department of Electrical
  Engineering and Computer Science at Northwestern University,
  Evanston, IL, 60208, USA.
    }
\thanks{
This work was supported in part by a gift from Futurewei Technologies and by the National Science Foundation under Grant No. CCF-1423040.
}
\small December 19, 2016}
\label{Joint}

\maketitle

\begin{abstract}
A scalable framework is developed to allocate radio resources across a large number of densely deployed small cells with given traffic statistics on a slow timescale. Joint user association and spectrum allocation is first formulated as a convex optimization problem by dividing the spectrum among all possible transmission patterns of active access points (APs). \BZ{To improve scalability} with the number of APs, \BZ{the problem is reformulated using local patterns} of interfering APs. To maintain global consistency among local patterns, inter-cluster interaction is characterized as hyper-edges in a hyper-graph with nodes corresponding to neighborhoods of APs. A scalable solution is obtained by iteratively solving a convex optimization problem for bandwidth allocation with reduced complexity and constructing a global spectrum allocation using hyper-graph coloring. Numerical results demonstrate the proposed solution for a network with 100 APs and several hundred user \BZ{equipments}. For a given quality of service (QoS), the proposed scheme can increase the network capacity several fold compared to assigning each user to the strongest AP with full-spectrum reuse.

\end{abstract}
\section{Introduction}
\label{ch5_sec:Intro}
Demand for wireless data service has continued to accelerate, driven by the proliferation of advanced mobile devices and data-intensive applications such as video on demand and cloud computing. As link level transmission rates approach their fundamental limits, network level improvements have drawn increasing attention. As proposed in the Long Term Evolution-Advanced (LTE-A) standard, a large number of small cells (picos, femtos, relays, remote radio heads, and WiFi access points (APs)) will be deployed under the coverage of legacy macro-cells to boost network capacity per unit area~\cite{3GPP_TR-36.932,3GPP_TR-36.814}. Compared to current and previous generations of cellular networks, it is much more challenging to allocate physical resources efficiently in a heterogeneous network (HetNet) formed by densely deployed APs.

One challenge is user association, which is traditionally decided according to the maximum reference signal receive power (maxRSRP) rule, i.e., each user equipment (UE) is assigned to the AP with the maximum receive power. In the downlink, maxRSRP association may lead to \BZ{severe load imbalance} between macro and pico tiers due to their differences in transmit power, signal propagation, and cell coverage~\cite{zhuang2015topology}. A simple remedy is \textit{range extension}~\cite{KhaBhuJi10EW,DamMonYon11TransWC}, where a bias factor is added to the RSRP of small cells. More complicated user association schemes have been studied for network utility maximization~\cite{shen2014distributed,fooladivanda2013joint,HonLuo13TransSP,KuaSpeDro12VTC,LinYu13GLOBE,YeRonChe13TransWC}. Many such optimization problems are considered with orthogonal frequency-division multiple access (OFDMA). Binary variables are in general used to indicate the UE-AP associations on a set of subcarriers or resource blocks. The optimization problem is often a difficult non-convex mixed-integer program.

Inter-cell interference management is especially challenging for cell edge UEs. An effective means to mitigate interference is to orthogonalize the spectrum allocations across adjacent cells according to certain frequency reuse patterns. Fractional frequency reuse is often more efficient and has been introduced to guarantee high throughput of cell center UEs~\cite{lei2007novel}. Dynamic fractional frequency reuse was recently studied in~\cite{
stolyar2008self,
chang2009multicell, 
ali2009dynamic,
madan2010cell,
liao2014base
} for OFDMA networks. The problem is usually formulated as deciding the subcarrier assignment to individual UEs according to instantaneous channel state information (CSI) and service demand. Due to the overhead of CSI exchange, such multi-cell dynamic spectrum allocation is usually coordinated within an autonomous cluster of no more than a few APs.

In this paper, user association and spectrum allocation are jointly considered based on the slow timescale optimization framework proposed in~\cite{zhuang2014traffic,zhuang2015energy}. Resource allocation on the fast (milliseconds) timescale, e.g, scheduling, usually depends on instantaneous channel and traffic information. Collecting this information for hundreds of APs and thousands of UEs is currently infeasible on a fast (milliseconds) timescale. Therefore, a slow timescale is proposed to allow adequate time during each decision period for information exchange and to solve the optimization problem. \BZ{At the same time, using a timescale of seconds to minutes allows one to track macro channel and traffic variations caused by user migration, service initiation/termination, and slow fading.} (To get a sense of the CSI overhead, consider 100 APs each sending 30,000 parameters (16 bits each) to the central controller once every minute. The aggregate data rate is about 0.8 Mbps, which is quite small.) Resource allocation on such a slow timescale is to improve the average user quality of service (QoS) given the average CSI and traffic statistics. Of course dynamic scheduling \BZ{based on instantaneous CSI} remains at each individual AP to adapt its own resources on a fast timescale.

This study builds on the network model described in~\cite{zhuang2015energy}. Each slice of time-frequency resource is shared by a subset of APs. We shall refer to this subset of APs as the corresponding {\em pattern}. In a HetNet of $n$ APs, there are in total $2^n$ patterns, which correspond to all possible ways the APs can share the spectrum. Assuming backlogged traffic and constant transmit power spectral densities (PSDs), a particular pattern determines the average signal to interference plus noise ratio (SINR) and \BZ{hence the spectral efficiency} of each link from a serving AP to a UE. In principle, the spectrum is divided into $2^n$ segments of variable bandwidths (some of which can be zero). We consider highly flexible user association where each AP can further divide each of its $2^{n-1}$ allocated patterns into arbitrary non-overlapping pieces to serve all or any subset of UEs in the network. With $k$ UEs (or UE groups), a network utility optimization problem with $O(kn2^n)$ variables is formulated to cover all possible allocations. We can optimize any performance metric that depends on the aggregate service rates, e.g., the average packet delay~\cite{zhuang2014traffic} or the number of active APs (for energy savings)~\cite{zhuang2015energy,kuang2016energy}. It has been proved in~\cite{zhuang2014traffic,zhuang2015energy} that, as long as the utility function to be maximized is concave, there exists an optimal solution that activates at most $k$ out of the $2^n$ patterns. For relatively small networks (with about 10 APs), the proposed solution demonstrates substantial performance gain over existing user association and interference management techniques. However, the computational complexity becomes prohibitive when the number of APs exceeds 20.

In this paper, we utilize the fact that inter-cell interference is a local phenomenon in a large network to pursue a fully scalable solution. The number of variables is reduced from $O(kn2^n)$ to $O(kn)$ by limiting the patterns to cover only local neighborhoods within dominating interferers. Global coordination is then guaranteed by introducing inter-cluster constraints on the local patterns. To obtain a feasible allocation, a combination of centralized convex optimization and hypergraph coloring is used. In the numerical results, we use the proposed scalable approach to allocate spectrum and assign UEs in a network with 100 APs. The solution achieves approximately three times the network throughput compared with full-spectrum reuse combined with maxRSRP user association in scenarios of interest.


The rest of the paper is organized as follows. Related work is reviewed in Section~\ref{sec:ReWork}. The global optimization problem is formulated in Section~\ref{sec:OptProb}, and is then relaxed to provide a scalable formulation in Section~\ref{sec:Relax}. A coloring-based approach is developed in Section~\ref{sec:ScaSol} to yield a feasible solution given the previous relaxation. Numerical results are shown in Section~\ref{sec:NumRes}. Section~\ref{ch5_sec:Con} concludes the paper.

\section{Related Work}
\label{sec:ReWork}
User association has been extensively studied for code-division multiple access (CDMA) networks~\cite{YatHua95TransVT,Han95JSAC,RasLiuTas97CL,SarManGoo01JSAC}. Results there suggest that joint user association and power control can significantly improve the performance of a CDMA network. Many recent studies focus on system utility
maximization in OFDMA HetNets, which often requires solving non-convex integer programs.
Game theory has also been used to derive simple distributed scheduling policies (e.g.,~\cite{JiaParWal08WCNC}). Optimal linear precoder design and base station selection are considered for uplink HetNets in~\cite{HonLuo13TransSP}. The authors of~\cite{EtkParTse07JSAC} studied spectrum sharing by strategic operators in the unlicensed band. While each operator is free to transmit over the entire common spectrum  subject to the maximum power constraint, leading to the tragedy of commons,~\cite{EtkParTse07JSAC} characterized more favorable Nash equilibria of both a one shot game and a repeated game. In contrast to the slow-timescale setting here, the aforementioned studies focus on dynamically updating user and resource allocation on a relatively fast timescale, which depends on the instantaneous channel realizations.

The stochastic geometry framework way proposed to evaluate and optimize the expected system performance over random topologies and channel conditions~\cite{SinBacAnd14,LinYu13GLOBE}. The approach does not apply to the optimization of resource allocation to all possible interference patterns as considered in this paper.

In~\cite{kuang2016optimal}, a slow-timescale model similar to the model considered in this paper was proposed. User association and spectrum allocation are jointly optimized to maximize the sum rate under proportional fairness constraints. There are two major differences between~\cite{kuang2016optimal} and our approach here. First, we allow rather realistic stochastic traffic, whereas~\cite{kuang2016optimal} is limited to backlogged traffic and rate maximization. Second, the proposed algorithm in~\cite{kuang2016optimal} avoids exponential complexity by limiting to a small number of global patterns {\em a priori}. Here, the proposed scalable solution includes all possible patterns as candidates.

\section{System Model}
\label{sec:OptProb}
\begin{figure}[!t]
\centering
\includegraphics[width=0.8\columnwidth]{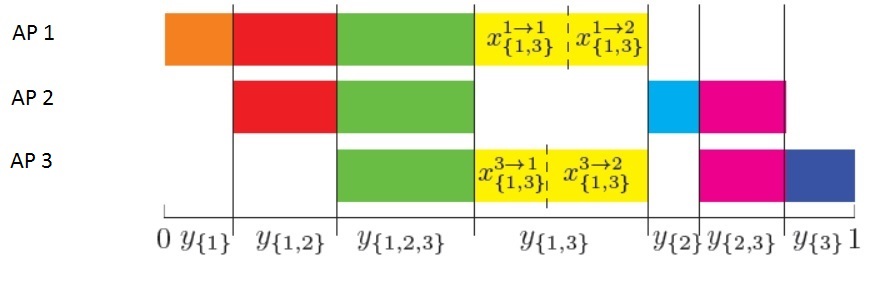}
\caption{Illustration of all patterns in a 3-AP network serving two UEs. (The allocations to the two users are revealed under pattern $\{1,3\}$.)}
\label{fig:3AP}
\end{figure}
We consider the problem of allocating a block of spectrum\footnote{The resource allocation is not limited to the frequency domain. Resources can be generally defined in the time-frequency plane.} of $W$ Hz across $n$ APs in the downlink of a HetNet. Denote the set of all \BZ{AP indices} as $\Ncal = \{1,\cdots,n\}$. We consider centralized global resource allocation on a slow timescale. The timescale is conceived to be on the level of a minute or even longer, which is in contrast to the millisecond frame-level scheduling in current LTE systems. The relatively long decision period \BZ{is more than sufficient for collecting} traffic and channel information from all APs, and also allows the use of advanced optimization tools to solve large optimization problems. On this slow timescale, we assume the spectrum is homogeneous \BZ{in the sense that all hertz are equally valuable}.

In this work, a {\em pattern} refers to a subset of APs, often denoted as $A\subset\Ncal$. There are $2^n$ distinct patterns in total. Every allocation can be viewed as a partition of the spectrum into segments (or colors) corresponding to those patterns. As illustrated in Fig.~\ref{fig:3AP} for the case of 3 APs, the spectrum is divided into 7 segments, excluding the empty pattern $\emptyset$. AP 1 has exclusive use of segment $\{1\}$, and shares segments $\{1,2\}$, $\{1,3\}$ and $\{1,2,3\}$ with the other APs. Let $y_A$ denote the bandwidth allocated to pattern $A$. Assuming the total available bandwidth is 1 unit, we have:
\begin{align}
\label{eq:TotalBW}
y_{\emptyset}+y_{\{1\}}+y_{\{2\}}+y_{\{3\}}+y_{\{1,2\}}+y_{\{1,3\}}+y_{\{2,3\}}+y_{\{1,2,3\}}=1.
\end{align}
An efficient allocation always sets $y_{\emptyset}=0$.

Each AP further divides each segment to serve different UEs. On the slow timescale, UEs near each other often have similar average channel conditions. To reduce complexity, it is then reasonable to treat UEs near each other with similar QoS requirements and propagation conditions as a group. The aggregate traffic of a UE group is modeled as a single queue. Denote the set of all $k$ groups of UEs as $\Kcal=\{1,\cdots,k\}$. This is without loss of generality, since in the extreme case each group contains a single UE. The packet arrivals for group $j$ UEs are modeled by a Poisson process with rate $\lambda^j$.

Let $x_A^{i\rightarrow j}$ denote the bandwidth used by AP $i$ to serve group $j$ under pattern $A$. If there are two UE groups in the entire system, then we must have:
\begin{align}
\label{eq:UECon}
x_A^{i\rightarrow 1} + x_A^{i\rightarrow 2} \leq y_A,\quad \forall A\subset \Ncal\; \text{and}\;i\in A.
\end{align}
As illustrated in Fig.~\ref{fig:3AP}, AP 1 divides $y_{\{1,3\}}$ (colored yellow) into two parts to serve the two groups, respectively, whereas AP 3 divides the same segment differently.

The spectral efficiency of link $i\rightarrow j$ (the link from AP $i$ to group $j$ UEs) over pattern $A$ is denoted by $s^{i\rightarrow j}_A$. Evidently, $s_A^{i\rightarrow j}=0$, if $i\not\in A$. Also, the exclusive spectrum has higher spectral efficiency than shared spectrum, e.g., $s^{1\rightarrow 1}_{\{1\}} > s^{1\rightarrow 1}_{\{1,2\}}$. In general,
\begin{align}
\label{eq:SE_ineq}
s^{i\rightarrow j}_A \geq s^{i\rightarrow j}_B,~\text{if}~A\subset B.
\end{align}
The spectral efficiency $s^{i\rightarrow j}_A$ can either be calculated based on pathloss and other impairments or measured over time. Let $\tau$ denote the average packet length in bits. We shall normalize the spectral efficiency by multiplying by $W/\tau$ so that the units are packets/second. For concreteness in obtaining numerical results, we use Shannon's formula to calculate spectral efficiency:
\begin{align}
\label{eq:SE}
&s^{i\rightarrow j}_A=\frac{W}{\tau}\,\mathbbm{1}(i\in A)\log_2\left(1+\frac{p_ig^{i\to j}}{\sum_{i'\in A\setminus\{i\}}p_{i'}g^{i'\to j}+n_j}\right) &~\text{packets/second}
\end{align}\noindent
where $\mathbbm{1}(i\in A)=1$ if $i\in A$ and $\mathbbm{1}(i\not\in A)=0$ otherwise, $p_i$ is the transmit PSD at AP $i$, $g^{i\to j}$ is the power gain of link
{$i\to j$}, and $n_j$ is the noise PSD at group $j$ UEs. Here we assume fixed flat transmit PSDs over the slow timescale. The link gain $g^{i\to j}$ includes pathloss and shadowing effects, again reflecting the slow timescale considered in this paper. Hence $g^{i\to j}$  and consequently $s^{i\to j}_A$ are constants in each decision period independent of the frequency.

The service rate to group $j$ contributed by AP $i$ over pattern $A$ is $s^{i\rightarrow j}_A x^{i\rightarrow j}_A$. The total service rate can be calculated by summing over all APs over all patterns. In the 3-AP example, all three APs may use parts of their assigned spectra to serve group 1, whose total service rate is then given by:
\begin{align}
\label{eq:rate1}
r^1 = s^{1\rightarrow 1}_{\{1\}}x^{1\rightarrow 1}_{\{1\}} + s^{1\rightarrow 1}_{\{1,2\}}x^{1\rightarrow 1}_{\{1,2\}} + s^{2\rightarrow 1}_{\{1,2\}}x^{2\rightarrow 1}_{\{1,2\}} + s^{1\rightarrow 1}_{\{1,2,3\}}x^{1\rightarrow 1}_{\{1,2,3\}} + \cdots + s^{3\rightarrow 1}_{\{3\}}x^{3\rightarrow 1}_{\{3\}}.
\end{align}\noindent
As shown later in the paper, each UE group is highly likely to be served by a single AP, and only a limited number of patterns are allocated nonzero bandwidth. The bandwidths also imply user association. In particular, $j$ is associated with AP $i$ if and only if $x^{i\rightarrow j}_A>0$ for some pattern $A$ with $i\in A$.


\subsection{Problem Formulation}
The spectral efficiencies are assumed known to the central controller. The spectrum allocation is defined by the variables $\bx =\left(x^{i\rightarrow j}_A\right)_{i\in \Ncal,\; j\in \Kcal,\; A\subset \Ncal}$ and $\by=\left(y_A\right)_{A\subset \Ncal}$, which determine the service rates of all UE groups $\br = [r^1,\cdots,r^k]$. Following~\cite{zhuang2015energy}, the joint user association and spectrum allocation problem is formulated similarly as~\ref{eq:ch5_Opt}:
\begin{varsubequations}{P0}
\label{eq:ch5_Opt}
\begin{align}
\underset{\br,\;\bx,\;\by}{\text{maximize}}~&~ u(r^1,\cdots,r^k)&\label{eq:Obj-Opt}\\
\text{subject to}~~&~r^j=\sum_{i\in \Ncal}\sum_{A\subset \Ncal} s^{i\rightarrow j}_A x^{i\rightarrow j}_A, &~j\in\Kcal \label{eq:Con1-Opt}\\
&~\sum_{j\in \Kcal} x^{i\rightarrow j}_A\leq y_A &~i\in\Ncal,\;A\subset \{1,\cdots,n\} \label{eq:Con2-Opt}\\
&~\sum_{A\subset \Ncal}y_{A}=1,&\label{eq:Con4-Opt}\\
&~ x^{i\rightarrow j}_A\geq 0,&~i\in\Ncal,\;j\in\Kcal,\;A\subset \{1,\cdots,n\} \label{eq:Con3-Opt}
\end{align}
\end{varsubequations}\noindent
where $u(\cdot)$ denotes the network utility function of interest. The constraints~\eqref{eq:Con1-Opt},~\eqref{eq:Con2-Opt} and~\eqref{eq:Con4-Opt} are the general forms of~\eqref{eq:rate1},~\eqref{eq:UECon} and~\eqref{eq:TotalBW}, respectively. \BZ{The decision variables to be optimized are $\bx$ and $\by$. In particular, $\by$ describes which (interference) patterns are activated and their allocated bandwidths. We refer to $\by$ as the {\em inter-cell allocation}. For given $\by$, $(x^{i\rightarrow j}_A)_{j\in\Kcal,A\subset\Ncal}$ denotes the bandwidths allocated across UE groups. We refer to $\bx$ as {\em intra-cell allocation}.} Due to the complexity of the model, we ignore the reduced interference during vacant periods of AP queues and assume all APs transmit at `conservative' rates obtained by assuming all queues are backlogged, as introduced in~\cite{zhuang2014traffic}. \BZ{Instead of the fixed transmit PSD in~\eqref{eq:SE}, we can apply a similar approach as in~\cite{zhuang2014traffic} to alternatively update bandwidth allocations and PSD distributions. However, the focus of this paper is to find a scalable solution to~\ref{eq:ch5_Opt}. Hence, fixed PSDs are assumed here.}

\BZ{\ref{eq:ch5_Opt} can be viewed as a generalization of previous collision-based formulations. Specifically, in a collision model, $s^{i\rightarrow j}_A=0$ if AP $i$ interferes with any other AP in set $A$; otherwise $s^{i\rightarrow j}_A =s^{i\rightarrow j}$, which is typically derived from the capacity of the link. Evidently, only links that do not interfere can be activated at the same time.~\ref{eq:ch5_Opt} then becomes a problem of optimizing the allocation over independent sets in Chapter 5 of~\cite[Ch. 5]{srikant2013communication}.}



The optimization problem~\ref{eq:ch5_Opt} is convex as long as the network utility function $u(r^1,\cdots,r^k)$ is concave in $\br$. Commonly used concave utilities include sum rate, minimum UE service rate (max-min fairness), and sum log-rate (proportional fairness). In this paper, we take average (negative) packet delay as the network utility function:
\begin{align}
\label{eq:AvgDelay}
u(r^1,\cdots,r^k)=-\sum_{j\in \Kcal}\frac{\lambda^j}{(r^j-\lambda^j)^+}
\end{align}\noindent
where $(x)^+$ equals $x$ if $x>0$ and $0$ otherwise. This utility function assumes exponential packet length with average $\tau$ bits/packet and the `conservative rate' defined in~\cite{zhuang2014traffic}. When considering the utility function~\eqref{eq:AvgDelay}, finite delay implies the stability condition:
\begin{align}
\label{eq:ConStab}
r^j > \lambda^j,~ j = 1,\cdots,k.
\end{align}

Although the number of patterns is exponential in $n$, it has been proved that there exists a globally optimal allocation with no more than $k$ nonzero patterns in~\cite{zhuang2015energy}. In a practical network, the average number of UEs per AP, $k/n$, is finite.
\begin{theorem}{(\cite{zhuang2015energy})}
\label{thm:Spec}
If the utility function in~\ref{eq:ch5_Opt} is concave, then there exists an optimal solution to~\ref{eq:ch5_Opt} with the following properties:
\begin{enumerate}
\item
The solution divides the spectrum into at most $k$ segments, i.e.,
\begin{equation}
\label{eq:ThmSpec}
\nonumber
\left|\{A\subset \Ncal~|~y_A>0\}\right|\leq k.
\end{equation}
\item
At most $n-1$ groups are jointly served by multiple APs, i.e.,
\begin{equation}
\label{eq:User}
\nonumber
\bigg|\left\{j~\bigg|~\exists i_1, i_2\in\Ncal, A_1, A_2\subset\Ncal,~\text{s.t.}, i_1\neq i_2, x^{i_1\to j}_{A_1}>0,~x^{i_2\to j}_{A_2}>0\right\}\bigg|\leq n-1.
\end{equation}
\item
The solution is throughput optimal, namely, it stabilizes all queues whenever there exists an allocation that can stabilize the queues.
\end{enumerate}
\end{theorem}

Property 2 follows the analogous proof of \emph{Proposition} 2 in~\cite{kuang2016optimal} using an argument based on characterizing the relation between UE groups and AP nodes as a bipartite graph. Property 3 is generalized from Theorem 3 in~\cite{zhuang2014traffic}.

\subsection{Numerical examples}
\label{ch5:OptNum}
\begin{figure*}
\centering
\subfloat[The same traffic distribution as in Fig.~\ref{fig:HeteroN10K46}.]{
\includegraphics[width = 0.48\columnwidth]{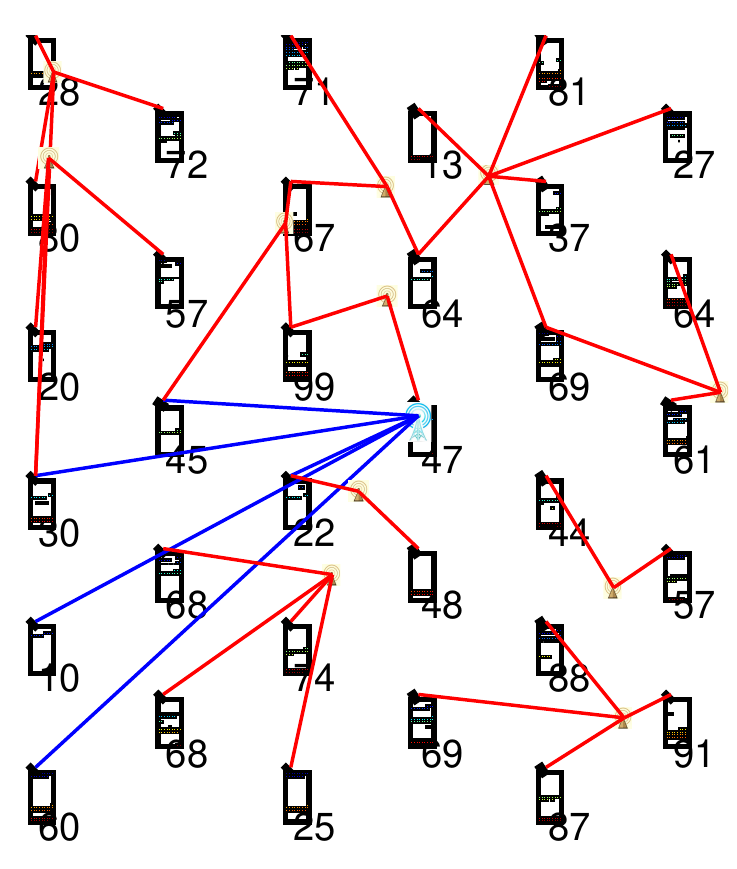}
\label{fig:SameDis}}
\subfloat[A different traffic distribution.]{
\includegraphics[width= 0.48\columnwidth]{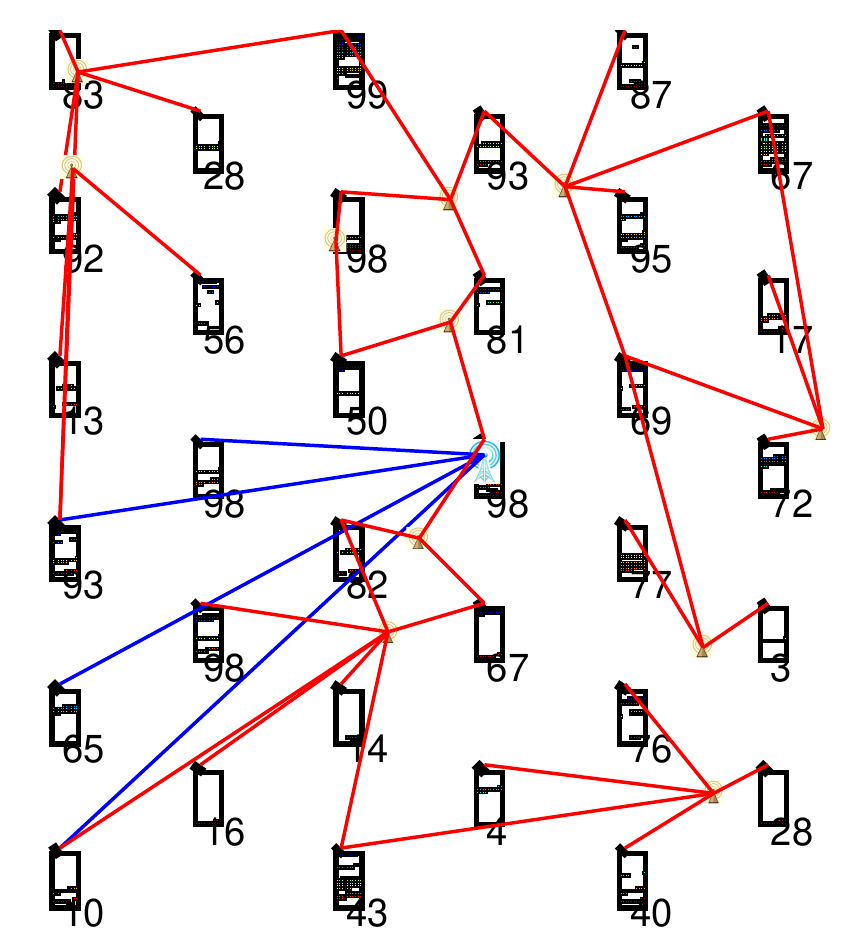}
\label{fig:DiffDis}}
\caption{Subcarrier allocation based on the solution obtained using Algorithm~\ref{alg:Overall} in a small HetNet with $N=12$ and $K=33$ at average traffic arrival rate of 1.33 packet/second for each user group.}
\label{fig:subcarrier}
\end{figure*}
We first illustrate the performance of the solutions to~\ref{eq:ch5_Opt} and its variations using a network cluster with $n=12$ APs and $k=33$ UE groups. The AP and UE group locations  over a $500\times 500~\text{m}^2$ area are depicted in Fig.~\ref{fig:subcarrier}. Common parameters used in simulations throughout the paper are given in Table~\ref{tab:par}.

\begin{table}[!t]
\caption{Parameter configurations.}
\label{tab:par}
\centering
\begin{tabular}{||c|c||}
\hline\hline
Parameter & Value/Function\\
\hline
pathloss exponent & 3\\
macro transmit PSD & 5 $\mu$W/Hz\\
pico transmit PSD & 1 $\mu$W/Hz\\
noise PSD & $1\times 10^{-7}$ $\mu$W/Hz\\
total bandwidth & 20 MHz\\
average packet length & 1 Mb\\
\hline\hline
\end{tabular}
\end{table}

\begin{figure}[!t]
\centering
\includegraphics[width=0.8\columnwidth]{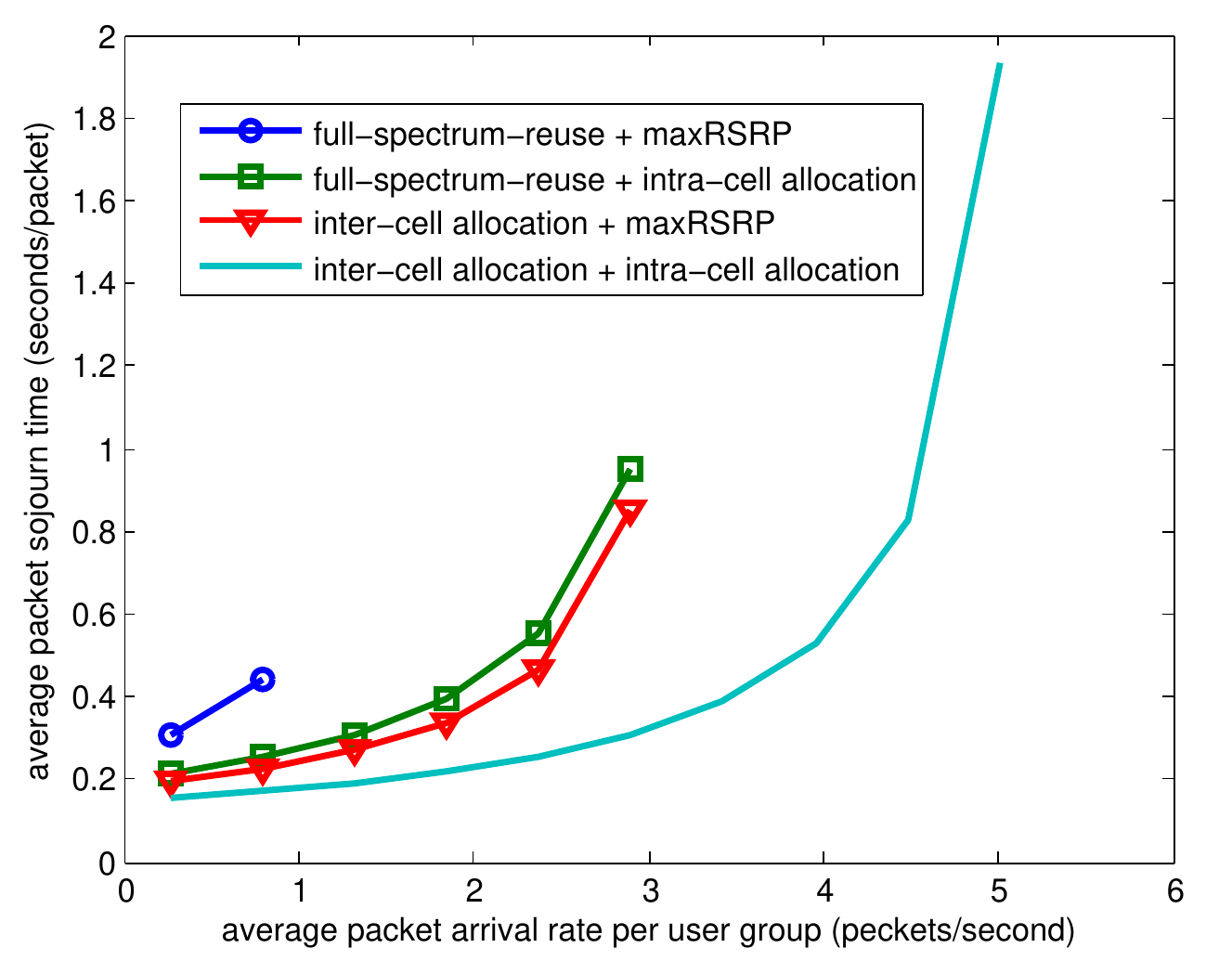}
\caption{Delay versus traffic intensity curves under a homogeneous setup.}
\label{fig:HomoN10K46}
\end{figure}

Fig.~\ref{fig:HomoN10K46} shows the delay versus traffic intensity performance for different allocation schemes. The cells are homogeneous pico cells. (All towers in Fig.~\ref{fig:subcarrier} represent pico APs for the simulation shown in Fig.~\ref{fig:HomoN10K46}.) The solution to~\ref{eq:ch5_Opt} is labeled as ``inter-cell allocation + intra-cell allocation". It is compared with three other schemes. The curve marked by circles corresponds to the full-spectrum reuse scheme with maxRSRP UE-AP association. In this case, $y_{\Ncal}=1$, and $y_A=0$ for all $A\neq \NN$. The curve with the square markers is obtained by optimizing the intra-cell allocation $\bx$ under full-spectrum reuse. The curve with the triangle markers is obtained by optimizing the inter-cell allocation $\by$ under maxRSRP association. The end of each curve in Fig.~\ref{fig:HomoN10K46} indicates the point where the system becomes unstable under the corresponding scheme. The maxRSRP association under full-spectrum reuse has the largest delay and rapidly becomes unstable as the traffic increases. Optimizing the intra-cell allocation alone provides marginal improvement, since it effectively optimizes only the user assignment without additional interference management. Optimizing the inter-cell spectrum allocation with maxRSRP association can be regarded as a more advanced version of enhanced inter-cell interference coordination (eICIC) in LTE-A on a slow timescale, which further reduces delay.\footnote{\BZ{eICIC uses the almost blank subframe (ABS) to mitigate interference in the time domain. Here optimizing $\by$ achieves the same function in the frequency domain. Certain APs (both macro and pico) are blanked on each pattern to reduce the interference to other APs. In contrast to the fixed length ABS, we allow an arbitrary fraction of the spectrum for each pattern.}} The optimal joint UE and spectrum allocation achieves the smallest delay and largest system throughput. In this homogeneous setup, \BZ{inter-cell allocation} plays an important role in mitigating interference among adjacent pico cells.

\begin{figure}[!t]
\centering
\includegraphics[width=0.8\columnwidth]{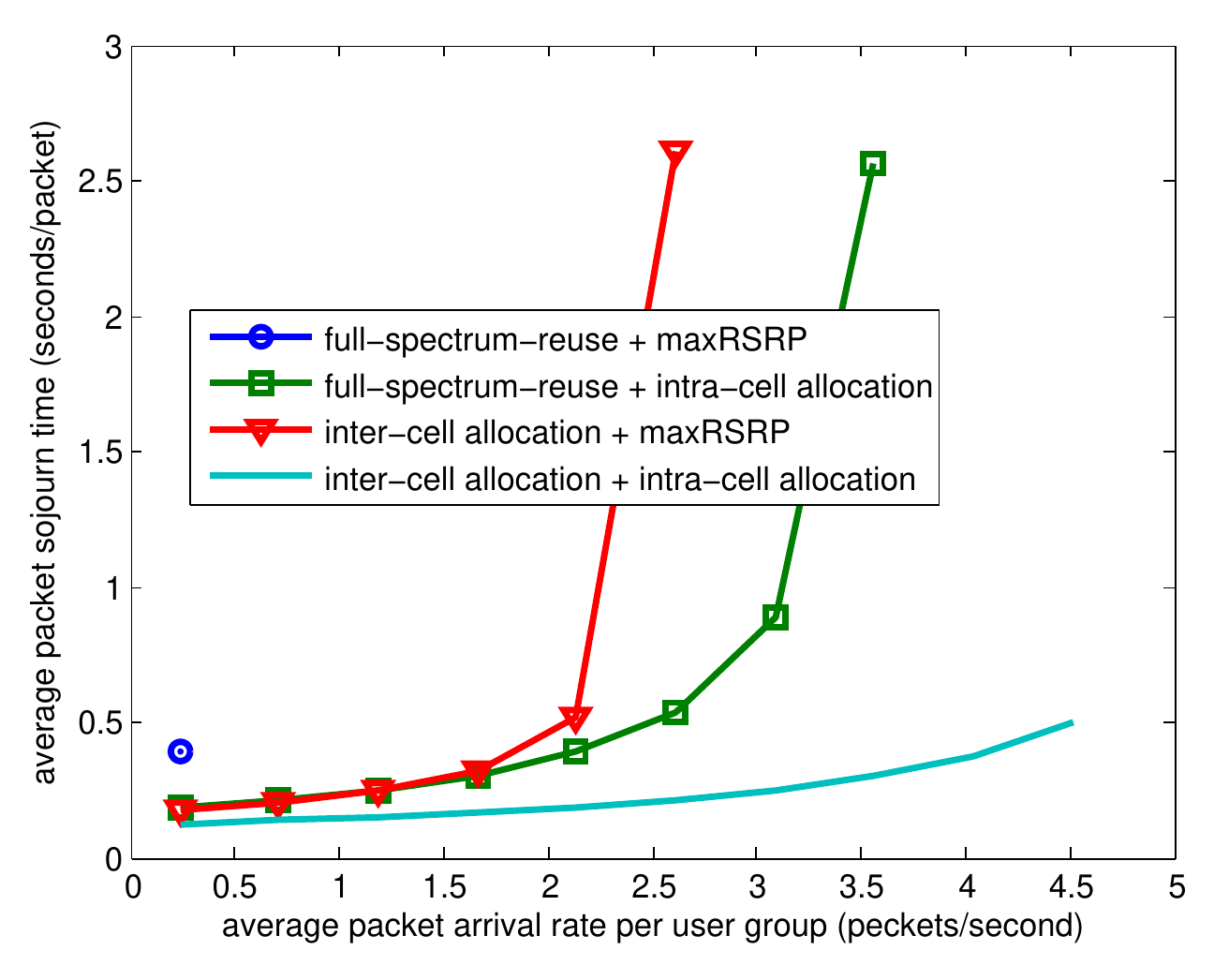}
\caption{Delay versus traffic intensity curves under a heterogenous setup.}
\label{fig:HeteroN10K46}
\end{figure}
The analogous delay versus traffic curves for a heterogenous setup are shown in Fig.~\ref{fig:HeteroN10K46}. There are one macro AP and eleven pico APs in the network, which are shown by the big and small towers in Fig.~\ref{fig:subcarrier}, respectively. Similar to the homogenous case, full-spectrum reuse with maxRSRP association suffers in both delay and throughput. Joint user association and spectrum allocation provides the maximum performance gain in both delay and throughput. However, optimizing user association under full-spectrum reuse shows significant delay and throughput improvements compared to optimizing the spectrum allocation with maxRSRP association \BZ{in the heavily loaded regime}. This is because most traffic will be directed to macro APs under maxRSRP association in this heterogeneous setup, which results in severely imbalanced traffic between macro and pico tiers \BZ{as traffic increases}.

\BZ{The optimal subcarrier allocation and user association shown in Fig.~\ref{fig:subcarrier} is for the same HetNet deployment used to generate Fig.~\ref{fig:HeteroN10K46}. The traffic load (average traffic arrival rate per UE group) is the same in Fig.~\ref{fig:SameDis} and Fig.~\ref{fig:DiffDis}. Fig.~\ref{fig:SameDis} has the same traffic distribution over different user groups as in Fig.~\ref{fig:HeteroN10K46}, while Fig.~\ref{fig:DiffDis} has a different traffic distribution.
The larger blue tower represents a macro AP, and smaller yellow towers represent pico APs. The 33 handsets represent 33 UE groups. The lines connecting different UE-AP pairs indicate the result association\footnote{A line connects AP $i$ and UE group $j$, if $\sum_{A\subset \NN}x^{i\rightarrow j}_A>0$.}. The colored tiles on the screen of each UE group represent the subcarriers that are used to serve it. The subcarrier allocation is a quantized version of the optimal solution to~\ref{eq:ch5_Opt} using 100 subcarriers. The number under each UE group (between zero and 100) represents the normalized traffic load of the corresponding group. The assignment algorithm achieves topology aware frequency reuse for interference management, as well as an efficient traffic aware spectrum allocation. Specifically, strongly interfering APs are assigned different subcarriers, and the same subcarriers are reused in cells that are far apart. The pico APs only serve adjacent UEs, while the macro APs tend to serve remote UEs in the coverage holes of the pico tier, which suggests an efficient traffic distribution between macro and pico tiers. We can see how the optimal allocation adapts to the traffic distribution change by comparing Figs.~\ref{fig:SameDis} and~\ref{fig:DiffDis}.}

\section{A Scalable Reformulation}
\label{sec:Relax}
\ref{eq:ch5_Opt} can be effectively solved using a standard convex optimization solver for networks with a small number of APs.
However, the number of variables in~\ref{eq:ch5_Opt} increases exponentially with the number of APs $n$, or more precisely, as $(kn+1)2^n+k$. The space and time complexities become prohibitive when the number of APs $n$ is large. This limitation is addressed by restricting to a fixed number of patterns selected {\em a priori} in~\cite{kuang2016optimal}. If all patterns are included, a large network would have to be divided into clusters of 10 to 20 APs for separate optimization. That could be far from optimal due to the inter-cluster interference on cluster boundaries. In this section, we develop a scalable relaxation of~\ref{eq:ch5_Opt} that is computationally viable for very large networks. How to obtain a good solution within the original feasible set of~\ref{eq:ch5_Opt} is presented in Section~\ref{sec:ScaSol}.


\subsection{Local patterns and allocations}
\label{subsec:Local}
Due to pathloss, radio signals effectively vanish beyond a finite coverage range. In a large wireless network, a UE group therefore only receives signals and interference from APs within a certain neighborhood. The data rate of the UE group therefore depends only on the spectra allocated to those APs. It would be sufficient to restrict the optimization of assigned spectrum to local patterns within the neighborhood, except that the allocation will have an impact on \BZ{overlapping neighborhoods, potentially cascading across the entire network}. The key idea here is to formulate the problem in terms of \emph{local} patterns in all neighborhoods, while maintaining the consistency of allocations between overlapping neighborhoods. To avoid underestimating interference, we assume all APs outside a neighborhood are backlogged (always transmitting) when determining the spectral efficiency of links within the neighborhood.

\begin{figure}[!t]
\centering
\includegraphics[width=0.5\columnwidth]{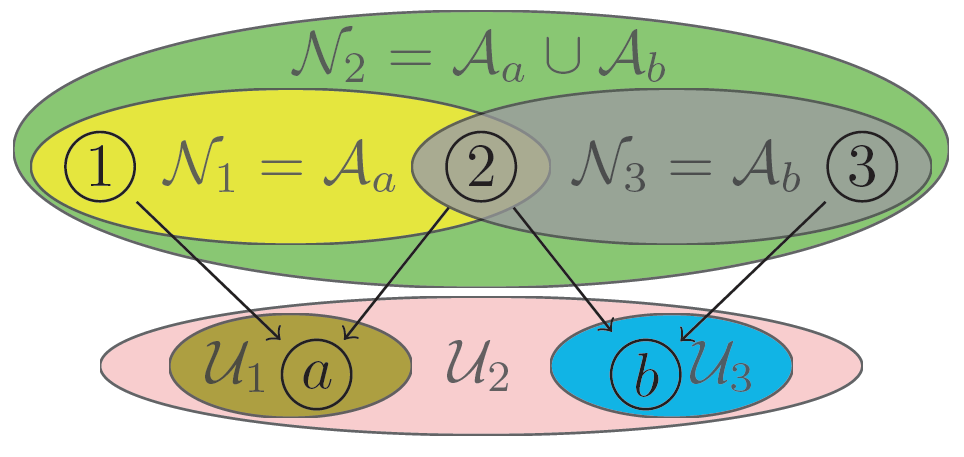}
\caption{Neighborhoods in the case of three APs and two UE groups.}
\label{fig:LocalPattern}
\end{figure}

Let $L$ denote the set of links with nonzero gain. Then $s^{i\rightarrow j}_A = 0$ for every link $(i\rightarrow j)\notin L$. We denote the neighborhood of  UE group $j$ as the set of APs that can reach group $j$:
\begin{align}
\label{eq:SetA}
\mathcal{A}_j=\{i|(i\rightarrow j)\in L\}.
\end{align}
On the AP side, denote the neighborhood of AP $i$ as the collection of UE groups AP $i$ can reach:
\begin{align}
\label{eq:SetU}
\mathcal{U}_i=\{j|(i\rightarrow j)\in L\}.
\end{align}
Finally, denote the set of all APs that interfere with AP $i$ (including $i$ itself) as $\mathcal{N}_i$:
\begin{align}
\label{eq:SetN}
\mathcal{N}_i = \cup_{j\in\mathcal{U}_i}\mathcal{A}_j.
\end{align}
An example with 3 APs and 2 UE groups is illustrated in Fig.~\ref{fig:LocalPattern}. \BZ{(The UE groups are denoted as $\{a,b\}$ to be distinguished from the AP indexes.)} We have $\NU_1=\{a\}$ (because AP 1 serves group $a$ only) and $\NA_a=\{1,2\}$ (UE group $a$ can only be served by APs 1 and 2). The neighborhood for AP 1 is $\NN_1=\{1,2\}$ since only AP 2 interferes with AP 1 at group $a$; while $\NN_2=\{1,2,3\}$, since AP 1 and 3 interfere with AP 2 at groups $a$ and $b$, respectively.

Given a reuse pattern $A$, the spectral efficiency of link $i\rightarrow j$, $s^{i\rightarrow j}_A$, only depends on APs in $A\cap \NA_j$. If we define the amount of spectrum assigned to link $i\rightarrow j$ in local cluster $B\subset\NN_i$ as $z^{i\rightarrow j}_B$, the summation in~\eqref{eq:Con1-Opt} can be rewritten as:
\begin{align}
\label{eq:LocalRate}
r_j &= \sum_{A\subset \NN}\sum_{i\in A}s^{i\rightarrow j}_A x^{i\rightarrow j}_A\nonumber\\
&=\sum_{i\in \NA_j}\sum_{B\subset \NN_i}s^{i\rightarrow j}_{B\cap\NA_j}z^{i\rightarrow j}_B.
\end{align}\noindent
The sum over
$B\subset \NN_i$ exhausts all patterns of transmitters that may interfere with AP $i$.

In what follows, we assume
\begin{align}
&|\mathcal{A}_j|\leq n_0,~ \forall j\in\Kcal \label{eq:Bound1}\\
&|\mathcal{U}_i|\leq k_0,~ \forall i\in\NN,\label{eq:Bound2}
\end{align}
where $k_0$ and $n_0$ are constants. That is, each UE group can be assigned to no more than $n_0$ APs (usually the strongest ones), and each AP can serve no more than $k_0$ UE groups (within its coverage). This implies an upper bound on the neighborhood sizes:
\begin{align}
\label{eq:size}
|\NN_i|\leq k_0n_0,~ \forall i\in\NN.
\end{align}\noindent
The total number of local variables $z_B^{i\rightarrow j}$ is then upper bounded by
\begin{align}
\label{eq:O(k)}
kn_02^{k_0n_0}=O(k),
\end{align}
which scales linearly with the network size. In contrast, the number of global variables $x_A^{i\rightarrow j}$ increases as $O(kn2^n)$.

\subsection{Local bandwidth constraint and consistency}
The set of global variables $\left(y_A\right)_{A\subset \NN}$ directly yields a global allocation. To develop a scalable solution, we replace them by $O(n)$ local variables $\left(y^i_B\right)_{i\in \NN, B\subset \NN_i}$, which specify the bandwidths of local patterns. Constraint~\eqref{eq:Con2-Opt} can be replaced by its equivalent representation using only local patterns:
\begin{align}
\label{eq:LocalBW}
\sum_{j\in\NU_i}z^{i\rightarrow j}_B\leq y^i_B.
\end{align}
Unfortunately, the global bandwidth constraint~\eqref{eq:Con4-Opt} has no direct equivalent representation using local variables.

To approximate solutions of~\ref{eq:ch5_Opt}, one approach is to replace~\eqref{eq:Con4-Opt} by the bandwidth constraints $\sum_{B\subset \NN_i}y^i_B\leq 1$ in each neighborhood $\NN_i$ and to add \textit{consistency constraints}:
\begin{align}
\label{eq:Consis}
\sum_{B\subset \NN_i:B\cap\NN_m=C}y^i_B=\sum_{B\subset \NN_m:B\cap\NN_i=C}y^m_B,~\forall i\in\NN,\;m\in\NN_i,\; C\subset \NN_i\cap\NN_m,\;C\neq\emptyset.
\end{align}
This holds for every non-empty local pattern $C$ used by two  overlapping neighborhoods $\NN_i$ and $\NN_m$. That is, they must assign the same total bandwidth to $C$ in each neighborhood. (In $\NN_i$, this is the sum over all $B$ that satisfies $B\cap\NN_m=C$.) There are $O(n)$ such constraints. \BZ{Consider $\Ncal_1$ corresponding to AP 1 and $\Ncal_3$ corresponding to AP 3 depicted in the 3-AP network in Fig~\ref{fig:3AP}. Since $\Ncal_1\cap \Ncal_3=\{2\}$, we must have:
\begin{align}
\label{eq:ExConsis}
y^1_{\{1,2\}}+y^1_{\{2\}}=y^3_{\{2,3\}}+y^3_{\{2\}}.
\end{align}
In this case,~\eqref{eq:ExConsis} states that the total bandwidths allocated to AP 2 is the same, regardless of the viewpoint.}

\subsection{Relaxed formulation}
The preceding replacements yield a relaxation of~\ref{eq:ch5_Opt} formulated as~\ref{eq:ch5_Opt3}:
\begin{varsubequations}{P1}
\label{eq:ch5_Opt3}
\begin{align}
\underset{\br,\;\by,\;\bz}{\text{maximize}}~&~u(r^1,\cdots,r^k
) &\label{eq:Obj-Opt3}\\
\text{subject to}~~&~ r^j= \sum_{i\in \NA_j}\sum_{B\subset \NN_i}s^{i\rightarrow j}_{B\cap\NA_j}z^{i\rightarrow j}_B, &~j\in\Kcal \label{eq:Con1-Opt3}\\
&\sum_{j\in\NU_i}z^{i\rightarrow j}_B\leq y^i_B, &~i\in\Ncal, \; B\subset\NN_i\label{eq:Con2-Opt3}\\
&\sum_{B\subset \NN_i}y^i_B\leq 1, & i\in\Ncal \label{eq:Con3-Opt3}\\
&\sum_{B\subset \NN_i:B\cap\NN_m=C}y^i_B=\sum_{B\subset \NN_m:B\cap\NN_i=C}y^m_B, & i\in\NN,\; m\in \NN_i,\nonumber\\
&& C\subset\NN_i\cap\NN_m,\;C\neq\emptyset \label{eq:Con4-Opt3}\\
&z^{i\rightarrow j}_B\geq0, &~i\in\Ncal, \; j\in \NU_i,\; B\subset\NN_i.\label{eq:Con5-Opt3}
\end{align}
\end{varsubequations}\noindent
Under assumptions~\eqref{eq:Bound1} and~\eqref{eq:Bound2}, the total number of variables and number of constraints in Problem~\ref{eq:ch5_Opt3} is $O(k+n)$.
Therefore, the convex optimization (assuming $u(r^1,\cdots,r^k)$ is concave) can be solved efficiently using standard convex optimization algorithms with computational complexity that is polynomial in $n$ and $k$.

\section{A Globally Feasible Solution}
\label{sec:ScaSol}
The relaxed problem~\ref{eq:ch5_Opt3} is {\em not} equivalent to~\ref{eq:ch5_Opt} since the constraints~\eqref{eq:Con3-Opt3} and~\eqref{eq:Con5-Opt3} do not imply the constraint on total bandwidth in~\eqref{eq:Con4-Opt}. Solving~\ref{eq:ch5_Opt3} gives a set of bandwidths to be assigned to local patterns across local neighborhoods. The challenge then is to relate these local assignments to global bandwidth assignments $A$ designated for patterns that encompass the entire network. We will refer to a particular global pattern $A$ as a {\em color}. We next present an algorithm for selecting a color assignment that satisfies the global bandwidth constraint~\eqref{eq:Con4-Opt} given a solution to~\ref{eq:ch5_Opt3}.


\subsection{A numerical example}
\begin{figure}[!t]
\centering
\includegraphics[width=0.5\columnwidth]{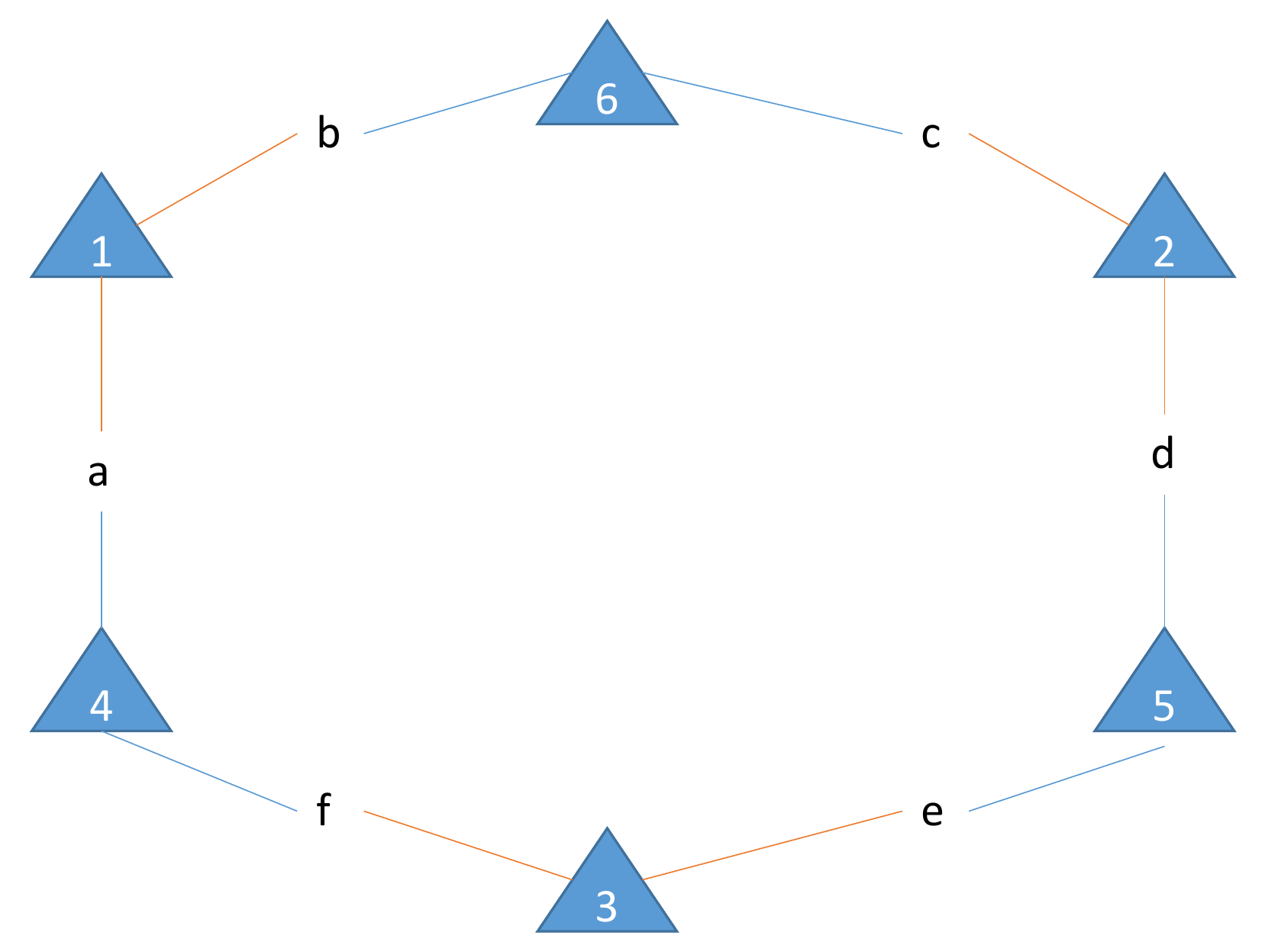}
\caption{An example network with $n=6$ and $k=6$.}
\label{fig:ExampleTopo}
\end{figure}
Fig.~\ref{fig:ExampleTopo} shows an example network with $\NN=\{1,2,3,4,5,6\}$ and  $\Kcal=\{a,b,c,d,e,f\}$. The links in the network with nonzero SNR are denoted by the lines. The neighborhoods in the network are given in Table~\ref{tab:Neighborhood}. Suppose the nonzero spectral efficiencies for each UE group are given in Table~\ref{tab:SpectralEff}. Let the packet arrival rate for each UE group be 20 packets/second.

\begin{figure}[!t]
\centering
\includegraphics[width=0.8\columnwidth]{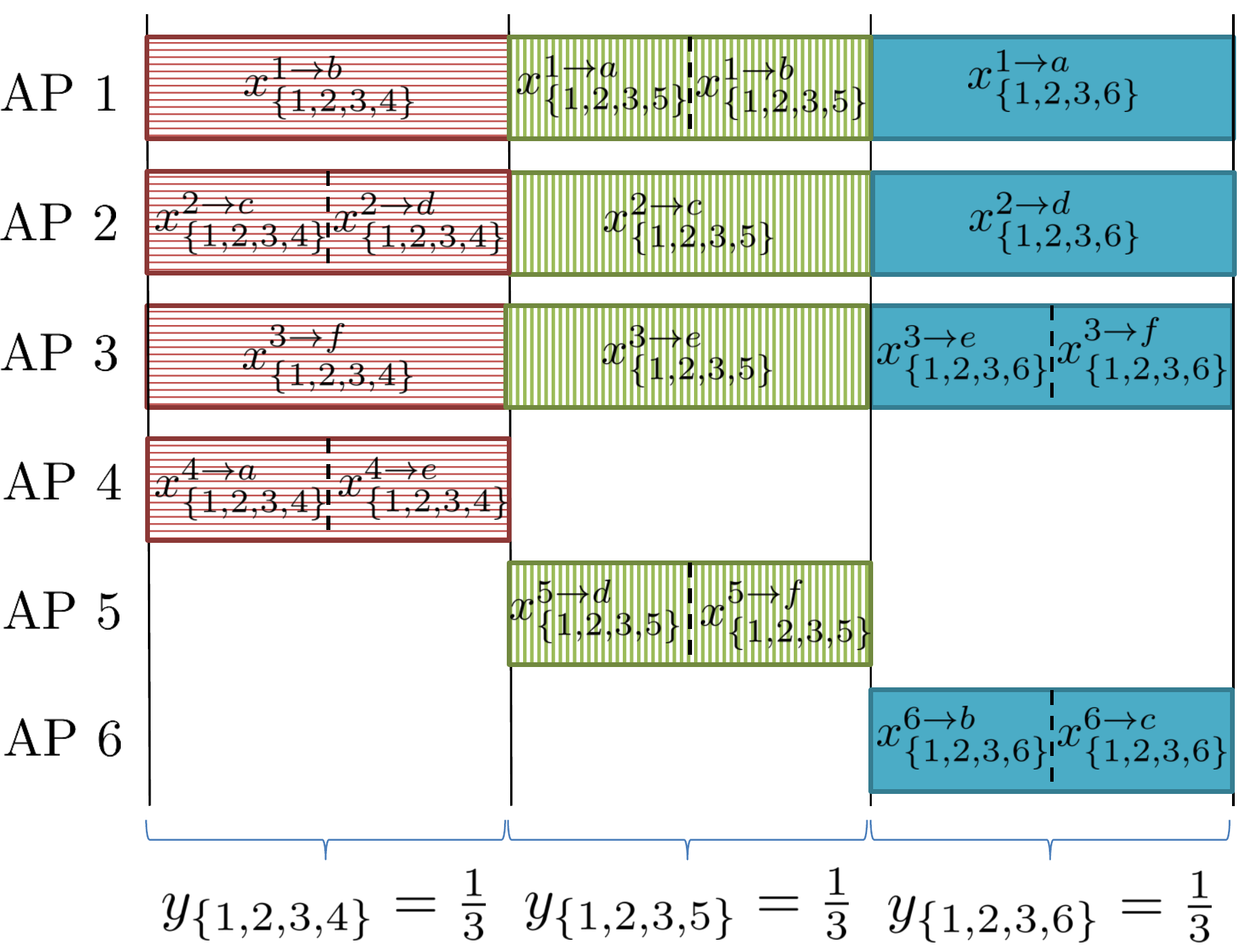}
\caption{The optimal solution for the network shown in Fig.~\ref{fig:ExampleTopo}}
\label{fig:OptAllocation}
\end{figure}

Let us now examine the optimal solution to~\ref{eq:ch5_Opt}. The corresponding spectral efficiencies of global patterns are obtained from the local spectral efficiencies in Table~\ref{tab:SpectralEff} by setting $s^i_{A}=s^i_{A\cap\NN_i},~ A\not\subset\NN_i$. The nonzero variables in the optimal solution are shown in Table~\ref{tab:OptGlo}. The minimum delay is 0.0331 seconds. The corresponding feasible spectrum allocation is shown in Fig.~\ref{fig:OptAllocation}. We can see the constraint~\eqref{eq:Con3-Opt3} is not binding at APs 4, 5, and 6.

\begin{table}
\caption{The optimal solution to~\ref{eq:ch5_Opt} for the network given by Fig.~\ref{fig:ExampleTopo}}.
\label{tab:OptGlo}
\centering
\begin{tabular}{||c|c|c|c|c|c|c||}
\hline\hline
$y_{\{1,2,3,4\}}=1/3$ & $x^{1\rightarrow b}_{\{1,2,3,4\}}=1/3$ & $x^{2\rightarrow c}_{\{1,2,3,4\}}=1/6$ & $x^{2\rightarrow d}_{\{1,2,3,4\}}=1/6$ & $x^{3\rightarrow f}_{\{1,2,3,4\}}=1/3$ & $x^{4\rightarrow a}_{\{1,2,3,4\}}=1/6$ & $x^{4\rightarrow e}_{\{1,2,3,4\}}=1/6$\\
$y_{\{1,2,3,5\}}=1/3$ & $x^{1\rightarrow a}_{\{1,2,3,5\}}=1/6$ & $x^{1\rightarrow b}_{\{1,2,3,5\}}=1/6$ & $x^{2\rightarrow c}_{\{1,2,3,5\}}=1/3$ & $x^{3\rightarrow e}_{\{1,2,3,5\}}=1/3$ & $x^{5\rightarrow d}_{\{1,2,3,5\}}=1/6$ & $x^{5\rightarrow f}_{\{1,2,3,5\}}=1/6$\\
$y_{\{1,2,3,6\}}=1/3$ & $x^{1\rightarrow a}_{\{1,2,3,6\}}=1/3$ & $x^{2\rightarrow d}_{\{1,2,3,6\}}=1/3$ & $x^{3\rightarrow e}_{\{1,2,3,6\}}=1/6$ & $x^{3\rightarrow f}_{\{1,2,3,6\}}=1/6$ & $x^{6\rightarrow b}_{\{1,2,3,6\}}=1/6$ & $x^{6\rightarrow c}_{\{1,2,3,6\}}=1/6$\\
\hline\hline
\end{tabular}
\end{table}

\begin{figure}[!t]
\centering
\includegraphics[width=0.5\columnwidth]{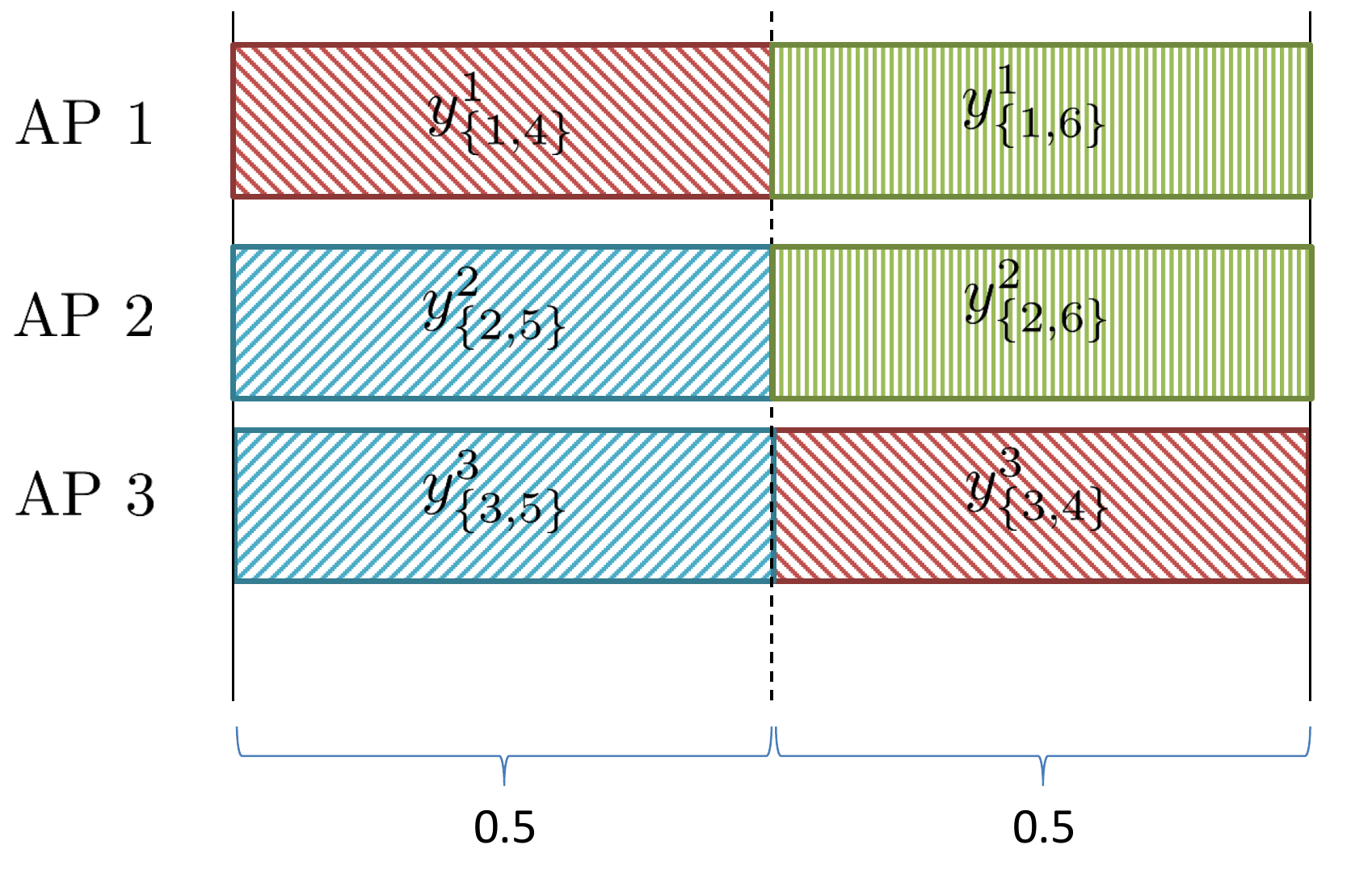}
\caption{Infeasible allocation corresponding to the solution to~\ref{eq:ch5_Opt3} given in Table~\ref{tab:OptSol}.}
\label{fig:Infeasible}
\end{figure}

Solving the corresponding~\ref{eq:ch5_Opt3} gives the optimal solution shown in Table~\ref{tab:OptSol}. The minimum delay is also 0.0331 seconds. According to Table~\ref{tab:OptSol}, exactly two patterns are active in each AP's local neighborhood. Also, APs 1, 2, and 3 use all available bandwidth, whereas APs 4, 5, and 6 each use half of the available bandwidth. This in fact violates the original constraints in~\ref{eq:ch5_Opt}. In other words, there exists no spectrum allocation that meets the local bandwidth requirements in Table~\ref{tab:OptSol}. To see this, we focus on the spectrum allocation at APs 1 to 3 as shown in Fig.~\ref{fig:Infeasible}. Those allocations are shown by the three rows from top to bottom. We start with AP 1, whose interference neighborhood is $\{1,4,6\}$. Of all possible patterns, only two are active: $y^1_{\{1,4\}}=y^1_{\{1,6\}}=0.5$. Without loss of generality, we assign the left half of the spectrum to APs 1 and 4 and the right half to APs 1 and 6. In AP 2's neighborhood $\NN_2=\{2,5,6\}$, the active patterns are $y^2_{\{2,5\}}=y^2_{\{2,6\}}=0.5$. To be consistent with the assignment in $\NN_1$, AP 2 has to share the right half of the spectrum with AP 6,\footnote{\BZ{Since AP 6 is only assigned half of the spectrum in the optimal solution, it has to use the same half of the spectrum in different clusters.}} and share the left half of the spectrum with AP 5. In AP 3's neighborhood $\NN_3=\{3,4,5\}$, the two active pattern are $y^3_{\{3,4\}}=y^3_{\{3,5\}}=0.5$. To be consistent with assignments in $\NN_2$, AP 3 and AP 5 must share the left half of the spectrum, forcing AP 3 and AP 4 to share the right half of the spectrum. However, this contradicts the assignments in $\NN_1$, and forces AP 4 to transmit over the entire spectrum, which is another contradiction.

\begin{table}[!t]
\caption{Local neighborhoods in the network shown in Fig.~\ref{fig:ExampleTopo}.}
\label{tab:Neighborhood}
\centering
\begin{tabular}{||c|c|c||}
\hline\hline
UE neighborhoods & AP neighborhoods & interference neighborhoods\\
\hline
$\mathcal{A}_a=\{1,4\}$ & $\mathcal{U}_1=\{a,b\}$ & $\NN_1=\{1,4,6\}$\\
$\mathcal{A}_b=\{1,6\}$ & $\mathcal{U}_2=\{c,d\}$ & $\NN_2=\{2,5,6\}$\\
$\mathcal{A}_c=\{2,6\}$ & $\mathcal{U}_3=\{e,f\}$ & $\NN_3=\{3,4,5\}$\\
$\mathcal{A}_d=\{2,5\}$ & $\mathcal{U}_4=\{a,f\}$ & $\NN_4=\{1,3,4\}$\\
$\mathcal{A}_e=\{3,5\}$ & $\mathcal{U}_5=\{d,e\}$ & $\NN_5=\{2,3,5\}$\\
$\mathcal{A}_f=\{3,4\}$ & $\mathcal{U}_6=\{b,c\}$ & $\NN_6=\{1,2,6\}$\\
\hline\hline
\end{tabular}
\end{table}

\begin{table}[!t]
\caption{Nonzero spectral efficiencies under local patterns of in the network shown by Fig.~\ref{fig:ExampleTopo}.}
\label{tab:SpectralEff}
\centering
\begin{tabular}{||c|c|c|c|c||}
\hline\hline
UE group $a$ & $s^{1\rightarrow a}_{\{1\}} = 100$ & $s^{4\rightarrow a}_{\{4\}}=2$ & $s^{1\rightarrow a}_{\{1,4\}}=5$ & $s^{4\rightarrow a}_{\{1,4\}}=1$\\
UE group $b$ & $s^{1\rightarrow b}_{\{1\}} = 100$ & $s^{6\rightarrow b}_{\{6\}}=2$ & $s^{1\rightarrow b}_{\{1,6\}}=5$ & $s^{6\rightarrow b}_{\{1,6\}}=1$\\
UE group $c$ & $s^{2\rightarrow c}_{\{2\}} = 100$ & $s^{6\rightarrow c}_{\{6\}}=2$ & $s^{2\rightarrow c}_{\{2,6\}}=5$ & $s^{6\rightarrow c}_{\{2,6\}}=1$\\
UE group $d$ & $s^{2\rightarrow d}_{\{2\}} = 100$ & $s^{5\rightarrow d}_{\{5\}}=2$ & $s^{2\rightarrow d}_{\{2,5\}}=5$ & $s^{5\rightarrow d}_{\{2,5\}}=1$\\
UE group $e$ & $s^{3\rightarrow e}_{\{3\}} = 100$ & $s^{5\rightarrow e}_{\{5\}}=2$ & $s^{3\rightarrow e}_{\{3,5\}}=5$ & $s^{5\rightarrow e}_{\{3,5\}}=1$\\
UE group $f$ & $s^{3\rightarrow f}_{\{3\}} = 100$ & $s^{4\rightarrow f}_{\{4\}}=2$ & $s^{3\rightarrow f}_{\{3,4\}}=5$ & $s^{4\rightarrow f}_{\{3,4\}}=1$\\
\hline\hline
\end{tabular}
\end{table}
\begin{table}[!t]
\caption{Nonzero $z^{i\rightarrow j}_B$ and $y^i_B$ in the optimal solution to~\ref{eq:ch5_Opt3} for the network shown in Fig~\ref{fig:ExampleTopo}.}
\label{tab:OptSol}
\centering
\begin{tabular}{||c|c|c|c||}
\hline\hline
$z^{1\rightarrow a}_{\{1,6\}}=0.5$ & $z^{1\rightarrow b}_{\{1,4\}}=0.5$ & $y^1_{\{1,4\}}=0.5$ & $y^1_{\{1,6\}}=0.5$\\
$z^{2\rightarrow c}_{\{2,5\}}=0.5$ & $z^{2\rightarrow d}_{\{2,6\}}=0.5$ & $y^2_{\{2,5\}}=0.5$ & $y^2_{\{2,6\}}=0.5$\\
$z^{3\rightarrow e}_{\{3,4\}}=0.5$ & $z^{3\rightarrow f}_{\{3,5\}}=0.5$ & $y^3_{\{3,4\}}=0.5$ & $y^3_{\{3,5\}}=0.5$\\
$z^{4\rightarrow a}_{\{1,3,4\}}=0.25$ & $z^{4\rightarrow f}_{\{1,3,4\}}=0.25$ & $y^4_{\{1,3\}}=0.5$ & $y^4_{\{1,3,4\}}=0.5$\\
$z^{5\rightarrow d}_{\{2,3,5\}}=0.25$ & $z^{5\rightarrow e}_{\{2,3,5\}}=0.25$ & $y^5_{\{2,3\}}=0.5$ & $y^5_{\{2,3,5\}}=0.5$\\
$z^{6\rightarrow b}_{\{1,2,6\}}=0.25$ & $z^{6\rightarrow c}_{\{1,2,6\}}=0.25$ & $y^6_{\{1,2\}}=0.5$ & $y^6_{\{1,2,6\}}=0.5$\\
\hline\hline
\end{tabular}
\end{table}

\subsection{The Coloring problem for subcarrier assignment}
\begin{figure}[!t]
\centering
\includegraphics[width=0.8\columnwidth]{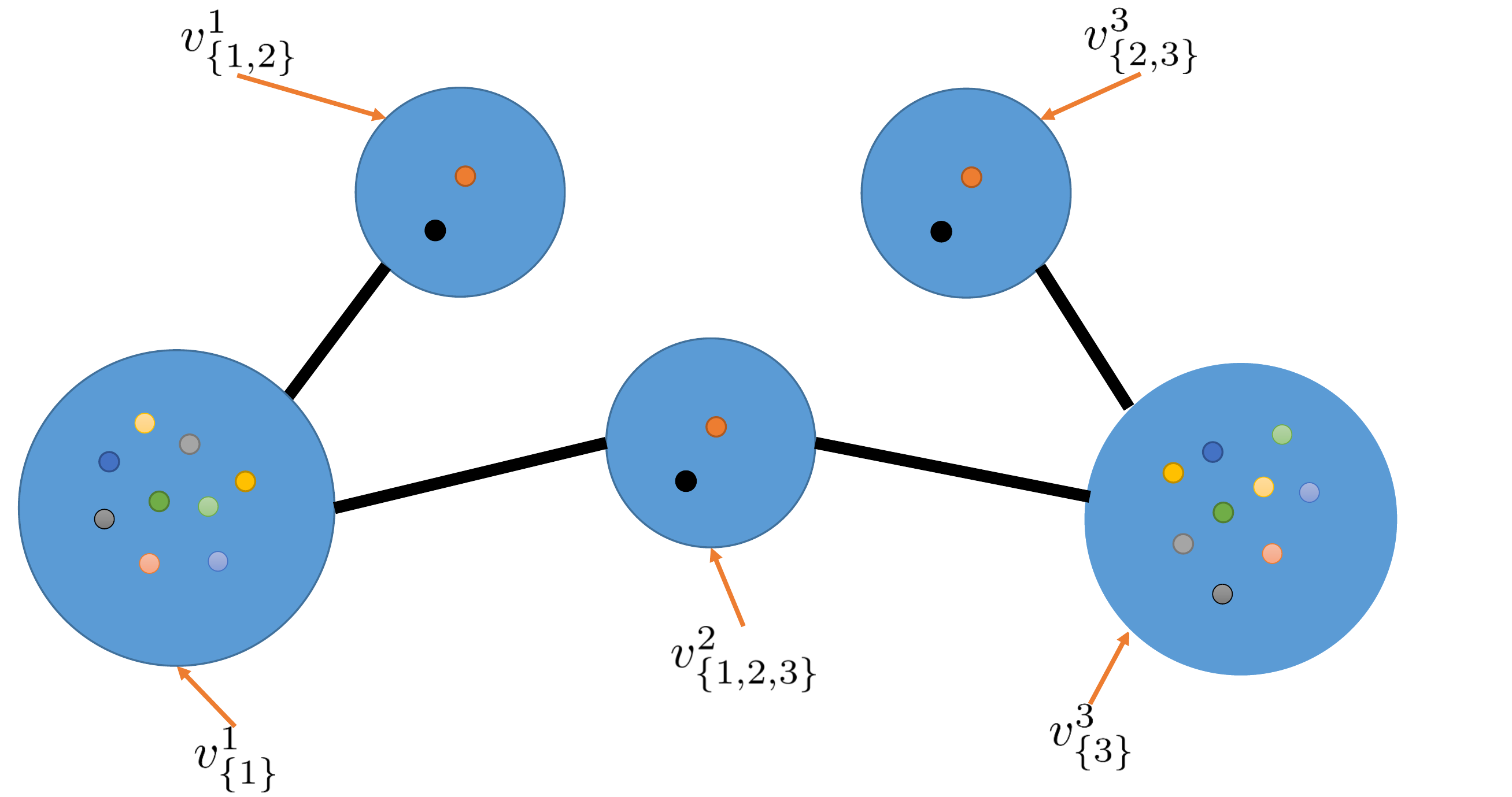}
\caption{An example of the hypergraph for the 3-AP example.}
\label{fig:HyperGraph}
\end{figure}

To address the feasibility issue, we formulate a discrete coloring problem based on the optimal (continuous) solution to~\ref{eq:ch5_Opt3}. Let the spectrum be divided into $n_s$ subcarriers of equal bandwidth. We first quantize the solution to~\ref{eq:ch5_Opt3}  as $\tilde{z}^{i\rightarrow j}_B= \lceil z^{i\rightarrow j}_B n_s\rceil$,\footnote{$\lceil x\rceil$ denotes the smallest integer that is greater than $x$.} which can be interpreted as the number of subcarriers or resource blocks (RBs) assigned to link $i
\rightarrow j$ over local pattern $B$.
The connection among all local patterns can be represented by a hypergraph $G\left(V,E\right)$ as shown in Fig.~\ref{fig:HyperGraph}. Here $V$ and $E$ are sets of vertices and edges, respectively. Each hypernode $v^i_B$ contains all vertices used by AP $i$ over local pattern $B$, i.e., $\{\tilde{z}^{i\rightarrow j}_B\}_{j\in\mathcal{U}_i}$. Each vertex within $v_B^i$ represents a unit of resource (subcarrier/RB) used by some link $i\rightarrow j$ over a local pattern $B$. The number of different subcarriers required by hypernode $v^i_B$ is $\sum_{j\in\mathcal{U}_i}\tilde{z}^{i\rightarrow j}_B$.

We use the network with three APs and two UEs depicted in Fig.~\ref{fig:LocalPattern} to show an example hypergraph. Let the spectral efficiencies of viable links be: $s^{1\rightarrow a}_{\{1\}}=6$, $s^{1\rightarrow a}_{\{1,2\}}=4$, $s^{2\rightarrow a}_{\{2\}}=5$, $s^{2\rightarrow a}_{\{1,2\}}=4$, $s^{2\rightarrow b}_{\{2\}}=2$, $s^{2\rightarrow b}_{\{2,3\}}=1$, $s^{3\rightarrow b}_{\{3\}}=4$ and $s^{3\rightarrow b}_{\{2,3\}}=1$. By solving~\ref{eq:ch5_Opt3} with $\lambda_a=4$ and $\lambda_b=2$, the nonzero variables in the optimal solution are: $z^{1\rightarrow a}_{\{1\}}=0.8856$, $z^{1\rightarrow a}_{\{1,2\}}=0.1144$, $z^{2\rightarrow a}_{\{1,2,3\}}=0.1144$, $z^{3\rightarrow b}_{\{3\}}=0.8856$, and $z^{3\rightarrow b}_{\{2,3\}}=0.1144$. Assuming $n_s=10$, we have $\tilde{z}^{1\rightarrow a}_{\{1\}}=9$, $\tilde{z}^{1\rightarrow a}_{\{1,2\}}=2$, $\tilde{z}^{2\rightarrow a}_{\{1,2,3\}}=2$, $\tilde{z}^{3\rightarrow b}_{\{3\}}=9$ and $\tilde{z}^{3\rightarrow b}_{\{2,3\}}=2$. The corresponding hypergraph is shown in Fig.~\ref{fig:HyperGraph}. The same subcarriers can be used by vertices in $v^1_{\{1,2\}}$ and $v^2_{\{1,2,3\}}$, since $\{1,2,3\}\cap\NN_1=\{1,2\}$, i.e., pattern $\{1,2,3\}$ is equivalent to pattern $\{1,2\}$ from AP 1's point of view. The same subcarriers can also be used by the vertices in $v^1_{\{1\}}$ and $v^3_{\{3\}}$, because AP 1 and AP 3 do not interfere with each other. In contrast, $v^2_{\{1,2,3\}}$ and $v^3_{\{3\}}$ cannot share the same subcarriers, since the interference neighborhood of AP $3$, $\NN_3=\{2,3\}$, includes AP $2$. (AP $2$ cannot transmit on the local pattern $\{3\}$.) If two hypernodes have such a conflict, we connect them with a hyperedge. More precisely, a hyperedge\footnote{A hyperedge is defined as the union of the vertices it connects.} $\left[v^{i_1}_{B_1}\cup v^{i_2}_{B_2}\right]$ connects two hypernodes $v^{i_1}_{B_1}$ and $v^{i_2}_{B_2}$ if either $\left(\NN_{i_1}\setminus B_1\right)\cap B_2\neq\emptyset$ or $\left(\NN_{i_2}\setminus B_2\right)\cap B_1\neq\emptyset$, where $\left(\NN_{i_1}\setminus B_1\right)$ includes all APs that are prohibited to use pattern $B_1$. On the contrary, if $(\NN_{i_1}\setminus B_1)\cap B_2 = (\NN_{i_2}\setminus B_2)\cap B_1=\emptyset$, there is no hyperedge between them.

The objective is to find a feasible subcarrier allocation using as few subcarriers as possible, which is equivalent to the \textit{strong vertex coloring} problem on the hypergraph~\cite{agnarsson2005strong}. A \textit{strong vertex coloring} of a hypergraph assigns distinct colors to vertices contained in a common hyperedge. For example, in Fig.~\ref{fig:HyperGraph} the nine vertices in hypernode $v^1_{ \{1\}}$ and the two vertices in hypernode $v^2_{\{1,2,3\}}$ must be colored by 11 different colors, as they are all contained in hyperedge $\left[v^1_{\{1\}}\cup v^2_{\{1,2,3\}}\right]$. Based on the definition of our hypergraph, a strong vertex coloring corresponds to a feasible subcarrier assignment. The strong coloring problem is to find a strong vertex coloring of hypergraph $G$ with the least number of colors, which is called the {\em strong chromatic number}, $\chi_s(G)$. The strong coloring problem can be formulated as the following linear integer program:
\begin{varsubequations}{P2}
\label{eq:ch5_OptColor}
\begin{align}
\underset{I_h,\;I^i_{B,h}}{\text{minimize}}~&~\sum_{h=1}^{nn_s} I_h &\label{eq:Obj-Color}\\
\text{subject to}~&~\sum_{h=1}^{nn_s}I^i_{B,h}=\sum_{j\in\mathcal{U}_i}\tilde{z}^{i\rightarrow j}_B, &~ i\in\NN,\; B\subset\NN_i \label{eq:Con1-Color}\\
&~I^{i_1}_{B_1,h}+I^{i_2}_{B_2,h}\leq I_h,
&~\left[V^{i_1}_{B_1}\cup V^{i_2}_{B_2}\right]\in E,~h\in\{1,\cdots,nn_s\} &\label{eq:Con2-Color}\\
&~I^i_{B,h}\in\{0,1\},&~i\in\NN,\;B\subset\NN_i,~h\in\{1,\cdots,nn_s\}&\label{eq:Con3-Color}\\
&~I_h\in\{0,1\}, &~h\in\{1,\cdots,nn_s\},\label{eq:Con4-Color}
\end{align}
\end{varsubequations}\noindent
where the binary variable $I_h$ indicates whether color $h$ is used, and the binary variable $I^i_{B,h}$ indicates whether any vertex in hypernode $v^i_{B}$ is colored by color $h$. The objective function in~\eqref{eq:Obj-Color} is the total number of distinct colors used. (At most $nn_s$ colors are needed to color the entire hypergraph, which corresponds to $n_s$ subcarriers per AP.) The constraint~\eqref{eq:Con2-Color} guarantees that conflicting hypernodes do not use the same color. The strong coloring problem on a hypergraph can be viewed as a traditional vertex coloring problem on the clique graph of the hypergraph~\cite{brandstadt1999graph}, which is known to be NP-hard for a general clique graph.

\begin{algorithm}
\caption{The heuristic greedy coloring algorithm.}
\label{alg:Color}
\begin{algorithmic}[]
\INPUT{$\tilde{\bz}$, $n_s$}
\OUTPUT{$\big(x^i_{B,h}\big)_{i\in N,~B\subset\NN_i,~h=1,\cdots, nn_s}$} and  $[I_1,\cdots,I_{nn_s}]$
\Init{$x^i_{B,h}\leftarrow0,~ I_h\leftarrow0,P_h\leftarrow\emptyset,~Q_h\leftarrow\emptyset, ~~\forall i\in\NN,~\forall B\subset\NN_i,~h=1,\cdots, nn_s$.}
\For{$i=1\cdots n$}
    \For{$B\subset\NN_i$}
        \State $l\leftarrow 0$
        \For{$j\in\NU_i$}
            \State $t\leftarrow0$, $h\leftarrow l$
            \While{$t<\tilde{z}^{i\rightarrow j}_B$}
                \State $h\leftarrow h+1$
                \If{$B\cap Q_h=\emptyset$ and $(\NN_i\setminus B)\cap P_h=\emptyset$}
                \begin{align*}
                &x^i_{B,h}\leftarrow 1,~I_h\leftarrow 1,\\
                &P_h\leftarrow P_h\cup B,~Q_h\leftarrow Q_h\cup(\NN_i\setminus B),\\
                &l\leftarrow h,~t\leftarrow t+1.
                \end{align*}
                \EndIf
            \EndWhile
        \EndFor
    \EndFor
\EndFor
\end{algorithmic}
\end{algorithm}

Our goal is to find an approximate solution to~\ref{eq:ch5_OptColor}, which achieves close to the minimum number of colors and requires relatively little computation.
A heuristic algorithm for obtaining such a solution is shown in Algorithm~\ref{alg:Color}. The main idea is to assign subcarriers to each AP one by one. When assigning subcarriers to a specific hypernode $v^i_B$, we avoid using new subcarriers if at all possible. Denote $P_h$ as the set of APs to which subcarrier $h$ has been assigned and $Q_h$ as the set of APs that cannot use subcarrier $h$ (due to pre-assigned hypernodes). If there exists any assigned subcarrier, which is not used by any vertex $v^{i'}_{B'}$ that is connected with $v^i_{B}$, such a subcarrier will be assigned to $v^i_{B}$. Then, $P_h$ and $Q_h$ are updated accordingly. If there is no such pre-assigned subcarrier, a new subcarrier will be assigned to $v^i_{B}$.

Algorithm~\ref{alg:Color} is an online algorithm in the sense that it sequentially assigns a color to one hypernode at a time. When assigning a color to a particular hypernode, only the information about hyperedges that are connected to previously colored hypernodes and the current hypernode is revealed. Algorithm~\ref{alg:Color} is \textit{frugal} in the sense that it prevents introducing new colors unless necessary. Define the degree of a vertex in hypergraph $G$ as the number of hyperedges connecting to it. It is shown in~\cite{agnarsson2005strong} that any online frugal algorithm is $\Delta(G)$-competitive,\footnote{The algorithm colors the graph with at most $\Delta(G)\cdot\chi_s(G)$ colors.} where $\Delta(G)$ is the maximum degree of the hypernodes in hypergraph $H$. In the worst case, $\Delta(G)$ can be quite large in general. However, in the hyper graph $G\left(V,E\right)$ generated by the optimal solution to~\ref{eq:ch5_Opt3}, $\Delta(G)\leq nk$, since at most $k$ global patterns will be active, as stated in Theorem~\ref{thm:Spec}, and each active pattern has at most $n$ hypernodes. We index the APs according to their distances from the center of the network in ascending order to take the most advantage of the online coloring algorithm. The underlying heuristic is that consecutively considering adjacent (strongly interfering) APs tends to orthogonalize the associated subcarrier assignment.

\subsection{The combined solution}
Algorithm~\ref{alg:Color} may end up using more than $n_s$ subcarriers. Hence we iteratively update the continuous allocation $(\bx,\;\by,\;\bz)$ by solving~\ref{eq:ch5_Opt3} and the discrete subcarrier allocation by solving~\ref{eq:ch5_OptColor}. This process is summarized in Algorithm~\ref{alg:Overall}. In each iteration, we first solve~\ref{eq:ch5_Opt3} with the total bandwidth constraint at each neighborhood,~\eqref{eq:Con3-Opt3}, set as $c$ (initially $c=1$). After obtaining $\tilde{\bz}$, Algorithm~\ref{alg:Color} is used to obtain $\left(I^i_{B,h}\right)_{i\in\NN, ~B\subset\NN_i, ~h=1,\cdots nn_s}$ and $\left(I_h\right)_{h=1,\cdots,nn_s}$. The total number of subcarriers used is $T=\sum_{h=1}^{nn_s} I_h$. If $(1-\delta)n_s<T\leq n_s$ , the algorithm terminates and outputs the corresponding subcarrier allocation $\left(I^i_{B,h}\right)_{i\in\NN, ~B\subset\NN_i, ~h=1,\cdots nn_s}$. Otherwise, the total bandwidth constraint is updated by $c=cN_s/T$, and the iterations continue. This change in total bandwidth constraint can be viewed as projecting the infeasible solution back to the feasible region.\footnote{If the objective is delay minimization, the projected solution is not guaranteed to be feasible, i.e., constraint~\eqref{eq:ConStab} may not be satisfied, in which case the output of Algorithm~\ref{alg:Overall} is treated as infeasible.}

\begin{algorithm}
\caption{The unified procedure to obtain an approximate solution.}
\label{alg:Overall}
\begin{algorithmic}[]
\INPUT{$\boldsymbol{\lambda},~n_s,~\delta$}
\OUTPUT{$\tilde{\bz}$,~$\big(I^i_{B,h}\big)_{i\in\NN, ~B\subset\NN_i, ~h=1,\cdots nn_s}$}
\Init{$T\leftarrow (1+2\delta)n_s$ and $c=1$.}
\While{$T>n_s$ or $T<(1-\delta)n_s$}
\State Get $\tilde{\bz}\leftarrow Q_{n_s}(\bz)$ by solving~\ref{eq:ch5_Opt3};
\State Get $\big(I^i_{B,h}\big)_{i\in\NN, ~B\subset\NN_i, ~h=1,\cdots nn_s}$ and $T=\sum_{h=1}^{nn_s} I_h$ by solving~\ref{eq:ch5_OptColor};
\State $c\leftarrow 1-(1-cN_s/T)_+$.
\EndWhile
\end{algorithmic}
\end{algorithm}

\section{Numerical Results}
\label{sec:NumRes}
The common parameters throughout this section are provided in Table~\ref{tab:par}. In the simulations, the UE groups are determined based on geographical location. For better illustration, we let them form a regular lattice. The APs are randomly uniformly dropped in the entire region. To limit the size of local neighborhoods, we only allow the nearest four dominant APs around each user group to serve the group. Only local patterns are used in Algorithm~\ref{alg:Overall}, where dependence on out-of-neighborhood APs is removed by assuming those APs have backlogged traffic and are always interfering. Although only four nearest APs are included in each {\em UE} neighborhood, the size of each {\em AP} neighborhood is usually larger than four due to interfering with different APs at different UE groups.

%
%

\subsection{Performance Comparison}
\begin{figure}[!t]
\centering
\includegraphics[width = 0.8\columnwidth]{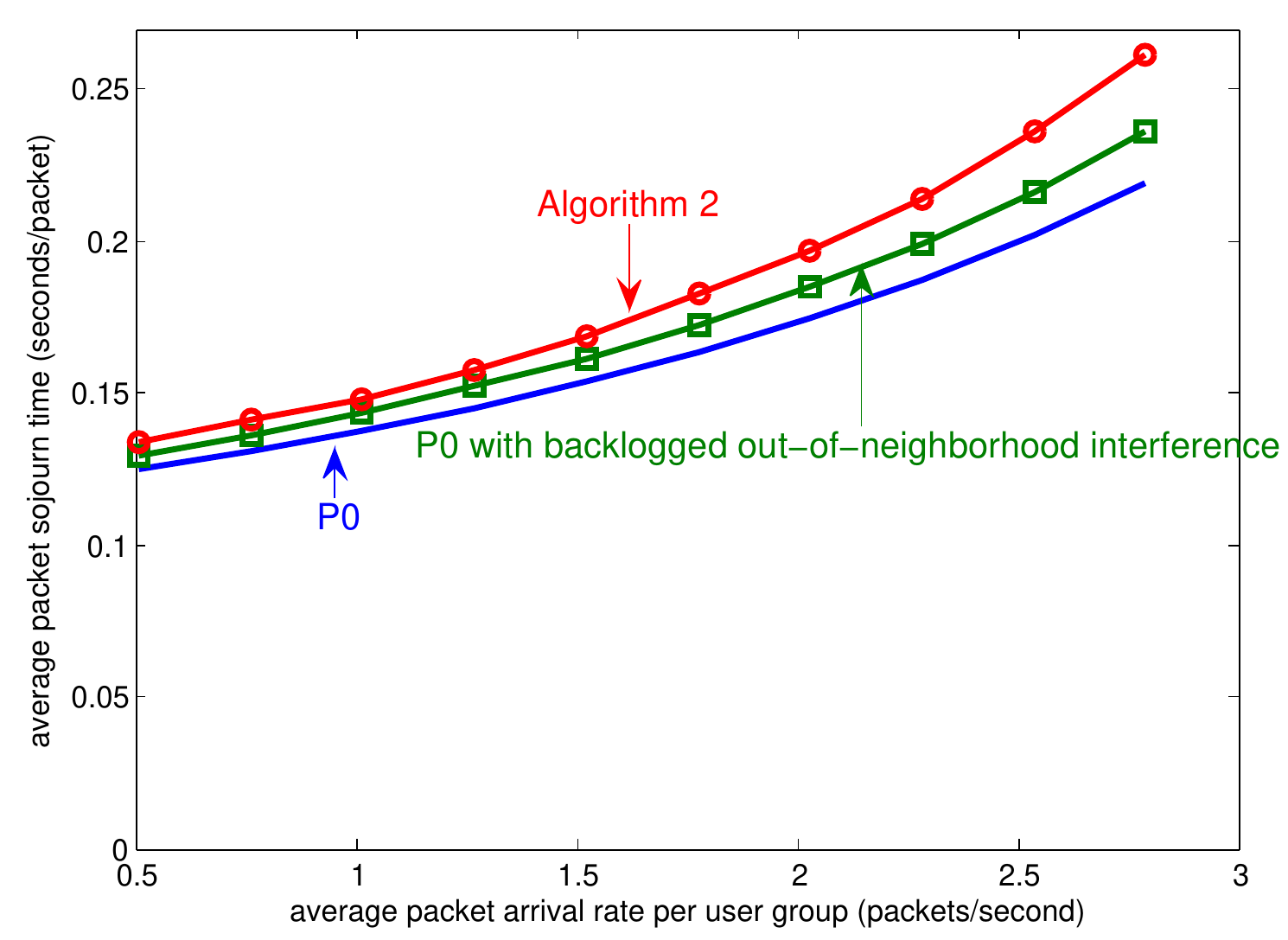}
\caption{Performance of the optimal assignment obtained by  solving~\ref{eq:ch5_Opt} with the scalable solution obtained from Algorithm~\ref{alg:Overall}.}
\label{fig:N25K126}
\end{figure}
We next compare the performance of the proposed scalable solution using Algorithm~\ref{alg:Overall} with the exact solution to~\ref{eq:ch5_Opt}. In order to solve~\ref{eq:ch5_Opt}, we use the same network with 12 APs and 33 groups as in Section~\ref{ch5:OptNum}. The bottom curve in Fig.~\ref{fig:N25K126} (with no marker) is obtained by solving~\ref{eq:ch5_Opt} directly. The top curve (with circle markers) is obtained using Algorithm~\ref{alg:Overall}. We can see they are relatively close for small to moderate traffic arrival rates. To understand the performance gap better, we also plot a middle curve (with square markers), which is the solution to~\ref{eq:ch5_Opt} with lower link spectral efficiencies by assuming APs outside each local interference neighborhood always transmit. This indicates that the performance loss of Algorithm~\ref{alg:Overall} is in part due to the assumed worst-case interference conditions and in part due to the suboptimal coloring scheme in each iteration.

\subsection{Performance in Large Networks}
\begin{figure}[!t]
\centering
\includegraphics[width = 0.8\columnwidth]{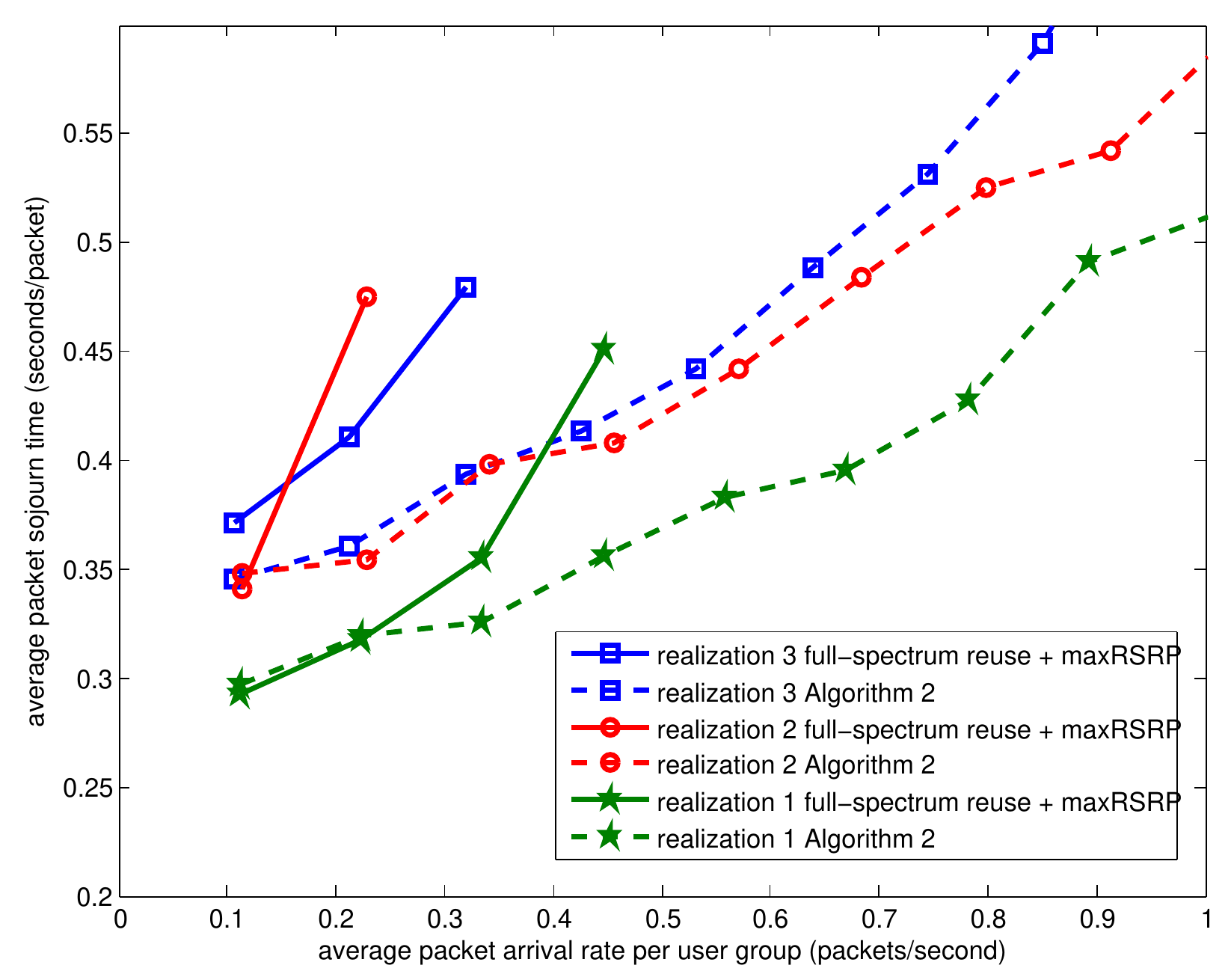}
\caption{Comparison of delay performance obtained from full-spectrum reuse with maxRSRP association and allocation obtained by applying Algorithm~\ref{alg:Overall} to a large size HetNet with $n=100$ and $k=314$.}
\label{fig:3Realization}
\end{figure}
\begin{figure}[h!t]
\centering
\includegraphics[width = 0.9\columnwidth]{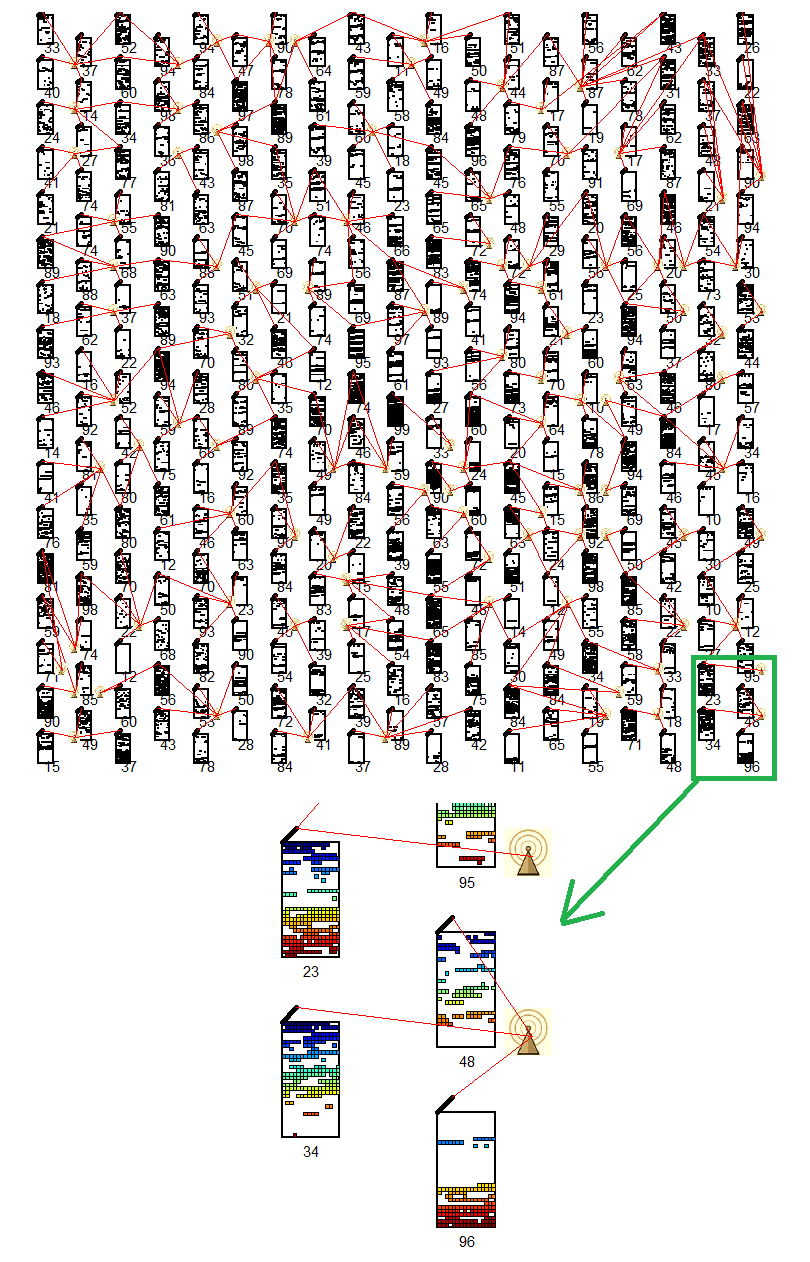}
\caption{Spectrum allocation and user association at 0.8 packets/second per user group for a large network with 100 APs and 314 UE groups.}
\label{fig:N100K314}
\end{figure}

In this section we show the performance of the proposed scalable solution in a large network with 100 small cells and 314 UE groups. Since directly solving~\ref{eq:ch5_Opt} is computationally infeasible, full-spectrum reuse with maxRSRP association is used as a reference. We assume $n_s=500$ subcarriers are available for the coloring algorithm. All other parameters are the same as in previous simulation results. In this simulation, we run Algorithm~\ref{alg:Overall} with random AP deployments and UE traffic distribution. The delay versus traffic intensity curves for three representative cases are shown in Figure~\ref{fig:3Realization}. The solid curves are for full-spectrum allocation with maxRSRP association, and the dashed ones are obtained using Algorithm~\ref{alg:Overall}. Each realization was obtained by dropping pico APs uniformly within the fixed area.

Each pair of solid and dashed curves with the same marker corresponds to the same random network realization. The proposed scalable solution achieves on average about three times the network capacity compared to the full-spectrum reuse allocation. In the very-low traffic regime, the proposed solution has similar (sometimes slightly worse) delay as the full-spectrum allocation. This is mainly due to the suboptimal projection at the end of each iteration. As the traffic load grows, the network quickly becomes unstable under the full-spectrum allocation. However, the proposed solution still achieves low delay and maintains stability. In the simulation, the threshold $\delta$ in Algorithm~\ref{alg:Overall} is chosen as 0.02. It is observed that the algorithm converges within two or three iterations for all realizations.

The obtained spectrum allocation and user association at average group packet arrival rate of 0.8 packets/second for the network realization 2 is shown in Fig.~\ref{fig:N100K314}. To clearly present subcarrier allocation, the local cluster located at the bottom right corner is shown in enlarged display. The coordination among APs is achieved through globally optimized spectrum allocation and user association to realize enhanced interference management and efficient network-wide load distribution.

\section{Conclusions}
\label{ch5_sec:Con}
We have considered joint user association and spectrum allocation in many cells over a slow timescale. The network utility maximization problem is formulated as a convex optimization over local neighborhoods with consistency constraints followed by subcarrier assignment using a coloring algorithm. Numerical results show substantial gains compared to full-spectrum reuse with maximum reference signal received power association.

The proposed solution iterates between bandwidth allocations across all possible reuse patterns and subcarrier assignments, and is scalable to large networks. The iterative algorithm finds an effective (although not optimal) solution to the original optimization problem for which the number of variables grows exponentially with the number of APs. Numerical examples demonstrate that the proposed scalable solution is capable of solving the global resource allocation problem for a HetNet with 100 APs. This appears to be the first attempt at obtaining a centralized near-optimal resource allocation for such a large network.

The proposed framework can potentially incorporate more sophisticated technologies such as coordinated multi-point (CoMP) transmission. Another possibility for further work is to find other scalable solutions that exploit the structure of the optimal solution.

\section*{Acknowledgement}
The authors thank Dr. Weimin Xiao and Dr. Jialing Liu for their valuable comments.

\end{document}